\documentclass[11pt]{article}
 \usepackage[a4paper,hmarginratio=1:1,vmarginratio=2:3,
totalwidth=15.6cm,  totalheight=21.cm]{geometry}

\usepackage{amsmath}
\usepackage{amssymb}
\usepackage{amsfonts}
\usepackage{epsfig}

\newcommand{\Vp}{{V_{2}}}
\newcommand{\Vm}{{V_{1}}}

\newcommand{\cD}{{\cal D}}

\newcommand{\cL}{{\cal L}}
\newcommand{\cM}{{\cal M}}

\newcommand{\cO}{{\cal O}}

\newcommand{\cZ}{{\cal Z}}
\newcommand{\cW}{{\cal W}}

\newcommand{\oD}{{\overline D}}

\newcommand{\ra}{\rightarrow}
\newcommand{\be}{\begin{equation}}
\newcommand{\ee}{\end{equation}}
\newcommand{\bea}{\begin{eqnarray}}
\newcommand{\eea}{\end{eqnarray}}

\DeclareMathSymbol{\mg}{\mathrel}{symbols}{"1D}

\newcounter{oldcounter}
\addtocounter{equation}{1}
\begin{document}
 \begin{flushright}
{CERN-PH-TH/2008-127, OUTP-0809P}\\
CPHT-RR036.0608,  LPT-ORSAY 08-56
\end{flushright}

\thispagestyle{empty}

\vspace{1cm}

\begin{center}
{\Large {\bf MSSM with Dimension-five Operators (MSSM$_5$)}}\vspace{1.cm}

  {\bf I. Antoniadis$^{\,a, b}$, E. Dudas$^{b, c}$,
   D.~M. Ghilencea$^{\,d}$, P. Tziveloglou$^{a,e}$ }\\
\vspace{0.5cm}
 {\it $^a $Department of Physics, CERN - Theory Division, 1211 Geneva 23,
 Switzerland.}\\[6pt]
 {\it $^b $CPHT, UMR du CNRS 7644,  \'Ecole Polytechnique, 91128
   Palaiseau Cedex, France.}\\[6pt]
 {\it $^c $LPT, UMR du CNRS 8627, B\^at 210, Universit\'e de Paris-Sud,
 91405 Orsay Cedex, France}
 \\[6pt]
 {\it $^d $Rudolf Peierls Centre for Theoretical Physics, University of
   Oxford,\\
 1 Keble Road, Oxford OX1 3NP, United Kingdom.}\\[6pt]
 {\it $^e $ Department of Physics, Cornell University, Ithaca, NY 14853 USA}
 \end{center}

\medskip

 \begin{abstract}
\noindent
We perform a general analysis of the R-parity conserving
dimension-five operators that can be present beyond the
Minimal Supersymmetric Standard Model. Not all these operators
are actually independent.
We present a  method  which employs  spurion-dependent
field  redefinitions that removes this ``redundancy'' and 
establishes the minimal,  irreducible  set of
these dimension-five operators.
Their potential effects on the MSSM Higgs sector are
discussed to show that the tree level bound $m_h\leq m_Z$
cannot be easily lifted within the approximations used, and
quantum corrections are still needed to satisfy the LEPII bound. An
ansatz  is provided for the structure of the remaining couplings in
the irreducible set of D=5 operators, which avoids
phenomenological constraints from flavor changing neutral currents.
  The minimal  set of operators  brings
new  couplings in the effective Lagrangian, notably ``wrong''-Higgs Yukawa
couplings and contact fermion-fermion-scalar-scalar interactions, whose
effects are expected to be larger than those generated in the MSSM at 
loop or even tree level. This has implications in particular 
for LHC searches for supersymmetry
by direct squark production.
\end{abstract}

\newpage

\setcounter{page}{1}
\tableofcontents{}

\section{Introduction}

The Standard Model (SM) and  its minimal supersymmetric version
(MSSM) are  thought to be the low energy limit of a more
fundamental theory valid at high scales (string theory,
extra dimensions, etc). In the absence of a detailed knowledge of this
theory (vacua degeneracy, moduli problem),
{\it effective} field theories provide a good framework on
searches for new physics. In such theories higher dimensional
operators are usually present.  They can be
 generated by compactification
or, in the case of 4D renormalisable theories,
by integrating out massive states of mass $M\gg m_Z$.
As a  result the low-energy effective
Lagrangian  below the scale $M$ contains a
set of  operators of dimension D$>$4.
The  effective field theory approach resides firstly
in organising these operators in a series of  powers of $1/M$.
In the leading order a smaller number of couplings
(parameters) are relevant and this leads to the possibility
of making low energy predictions, little dependent on the
details of the high scale theory (in many cases unknown anyway).
For practical purposes one can consider,
in addition to the SM or MSSM Lagrangian, the set of all higher
dimensional operators of a given dimension with some unknown
coefficients and investigate their implications  for
electroweak or TeV scale physics.
A second organising principle is that,
for a given order in $1/M$, one may use in addition symmetry
arguments inspired by phenomenology, to reduce further
the number of parameters.

When studying the effects of higher dimensional operators
one aspect is often overlooked. This refers to the fact that
in an effective field theory not all operators of a given
dimension (suppressed by a fixed power of $1/M$) are actually
independent. Within a given such set of  operators,
general field transformations allow one to eliminate
those operators which are redundant, and identify the minimal
irreducible set of independent operators.
The advantage of this result is that it simplifies considerably
the study of the models, by removing redundant couplings (parameters)
 of the theory. The purpose of this
work  is to show explicitly how one can identify
the minimal irreducible set of such operators for a
particular example.
We consider   the  MSSM\footnote{For a review
 see \cite{Nilles:1983ge}.} extended by all dimension-five
operators that conserve $R$-parity symmetry \cite{Farrar}
and we identify the
minimal  irreducible set of these. The method is general and can be
applied to other models, too.

Since supersymmetry is broken, the fields' transformations
should take into account effects of supersymmetry
breaking associated with the higher dimensional operators.
This is done by using  spurion-dependent transformations.
 Some operators are
``redundant'' in that they can be eliminated completely or
they only change/renormalise
the standard soft terms and supersymmetric $\mu$-term;
such operators can be ``gauged away''.
 In the new fields' basis the final number of parameters
 is reduced and  calculations and predictions for
physical observables  can be more easily made. We provide an ansatz for
the remaining couplings which allows one to avoid the effects of Flavor
Changing Neutral Currents (FCNC),
and reduces further the number of these couplings. One
consequence is the generation of new effective interactions 
in the Lagrangian of
the type (quark-quark-squark-squark) with potentially large effects
in squark production  compared to those generated in the MSSM. These
are largest for the top/stop quarks.
This  can be important for LHC supersymmetry
searches by direct squark production. Additional ``wrong''-Higgs
couplings, familiar in the MSSM at the loop level
\cite{Haber,M0,M1}, are also generated with a numerical coefficient
that can be  larger than the loop-generated MSSM one.
Again, these are largest for the top and also bottom sector at large
$\tan\beta$. We discuss some of the associated phenomenological
implications.

We  show that in the model discussed
the Higgs sector is simplified, despite the initial
presence of two D=5 operators and their associated
spurion dependence. The ``redundant'' operator
that can be removed by field redefinitions does not change the physics
of the Higgs sector. It also
turns out that for the MSSM lightest Higgs
the tree level bound  $m_h\leq m_Z$ is not easily lifted
by the D=5 operators (with one exception that we discuss).
The conclusion is that
in the approximation considered the MSSM Higgs sector is
rather stable under the addition of D=5 operators
and quantum
corrections are still needed to lift it above LEPII bound
\cite{higgsboundLEP}.
This conclusion  changes  if the massive states that
induce the D=5 operators in the first instance
are sufficiently light  not be integrated out
but  considered together with the other MSSM
states  when analysing their implications.

The plan of the paper is as follows. In the next Section we present
the general D=5 operators  that can be present beyond the MSSM,
preserving R-parity. We then identify the minimal, irreducible set of
these operators. Although of dimension-five, 
they can still induce too-large, dangerous FCNC effects, for
arbitrary coefficients. An ansatz avoiding this problem is presented,
together with its phenomenological implications, in
Section~\ref{ss1p}. In Section~\ref{ss2}, we analyse the effects on
the Higgs sector that D=5 operators can bring. We  show that these
cannot avoid the MSSM tree level upper bound on the lightest Higgs
($m_h\leq m_Z$), with one exception where a marginal
increase above $m_Z$ can be present.
 We check explicitly that, as
expected, an operator that belongs to the  redundant class cannot
change the upper bound on the lightest Higgs and only renormalizes
soft masses or the $\mu$ term. In Appendix~\ref{appendixA} and
 Appendix~\ref{appendixB} we show in detail how the higher
dimensional operators of the type discussed in the text occur at low
energies, by integrating out massive supermultiplets (that could be
present beyond MSSM \cite{Derendinger:1983bz}),
in the absence (Appendix~\ref{appendixA}) and in the
presence (Appendix~\ref{appendixB}) of gauge interactions.
Appendix~\ref{appendixC} identifies the most general supersymmetry
breaking terms that a particular type of D=5 operator  discussed in
Section~\ref{ss1} can bring. Finally Appendix~\ref{appendixD}
provides technical details of the calculation of the Higgs spectrum
discussed in the text.

\section{Higher dimensional operators: a general discussion.}
\label{ss1}

In this section we find the minimal, irreducible set of R-parity
 conserving
dimension-five operators that can be present beyond the MSSM.
Consider
\medskip
\be\label{L1}
\cL=\cL^{(4)}_{MSSM}+\cL^{(5)}
\ee
Here $\cL_{MSSM}^{(4)}$ is the
 standard R-parity conserving MSSM Lagrangian and $\cL^{(5)}$
is a Lagrangian of R-parity conserving dimension-five  operators,
to be introduced shortly.  Further

\bea\label{LMSSM}
\cL^{(4)}_{MSSM}
 &= &
\int d^4\theta \,\Big[\,
\cZ_1\,H_1^\dagger \,e^\Vm\,H_1+
\cZ_2\,H_2^\dagger \,e^\Vp\,H_2\Big]
 +\cL_K
\nonumber\\[8pt]
+ && \!\!\!\!  \!\!\!\!
\bigg\{\int d^2\theta\,\Big[-\,H_2 \,Q\,\lambda_U\,U^c
- Q\,\lambda_D\,D^c\,H_1
-  L\,\lambda_E\,E^c\,H_1+
\mu\,H_1\,H_2
\Big]+h.c.\bigg\}
\eea

\medskip\noindent
$\cL_{K}$ accounts for the kinetic terms of the quark and lepton superfields 
$Q,U^c,D^c,L,E^c$ and
for the gauge kinetic part, as well as for their associated soft
breaking terms obtained using spurion field formalism.
In the MSSM\footnote{$U^c, D^c, E^c$
denote anti-quark/lepton singlet chiral superfields
of components  $f_R^c\equiv (f^c)_L$ and $\tilde f_R^*$, $f=u,d,e$, while
$Q$ and $L$ denote the left-handed quark and lepton superfields doublets.}
$\Vm\equiv \,g_2\, V_W^i\, \sigma^i -g_1 \, V_Y$,  ($H_1$ has $Y_{H_1}=-1$)
and  $\Vp\equiv \,g_2\,V_W^i\,\sigma_i+g_1 \, V_Y$,
where $V_Y$ and $V_W$ are vector superfields of the $U_Y(1)$-hypercharge and
$SU(2)_L$ respectively, and $g_1$ and
$g_2$ are the corresponding couplings\footnote{We denote a product of two
$SU(2)$ doublets (columns)   $H_2\,Q\,\lambda_U\,U^c\!\equiv\!
H_2^T\,(i\sigma_2)\,Q\,\lambda_U\,U^c$
in a matrix notation, which helps us to avoid extra $SU(2)$ indices;
also $H_1\,H_2\equiv H_1^T\,(i\sigma_2)\,H_2$; similar
convention is used below.}.
Finally, $\lambda_F$, $F=U,D,E$ are $3\times 3$ matrices in the flavor
space. Note that
\bea
\cZ_i\equiv \cZ_i(S,S^\dagger),\qquad \lambda_F\equiv
\lambda_F(S),\,\,\,\,\,F:U,D,E,\qquad
\mu\equiv \mu(S)\qquad
\eea
where  $S \equiv M_s\,\theta^2$ is the spurion parametrising the soft
supersymmetry breaking and $M_s$ is the supersymmetry breaking scale.
In the following we use the notations
\medskip
\bea
&& \cZ_1 = 
1 + a_1 S +  a_1^* \,S^{\dagger} + a_2 S S^{\dagger} \ ,
\nonumber \\
&& \cZ_2 = 
1 + b_1 S +  b_1^* \,S^{\dagger} + b_2 S S^{\dagger}
\ .
\eea

\medskip\noindent
The complete set of dimension-five operators in MSSM, which preserve
R-parity is given by\footnote{
For a general discussion of D=5 operators with discrete symmetries
 see \cite{Ibanez:1991pr}.}
\medskip
\bea\label{ll5}
\cL^{(5)}&=&
\frac{1}{M}\bigg\{
\int d^2\theta
\,\,\Big[ Q\,U^c\,T_Q\,Q\,D^c+ Q\,U^c\,T_L\,L\,E^c+ \lambda_H
 (H_1 H_2)^2\,\Big]+h.c.\bigg\}
\nonumber\\[8pt]
&+&
\frac{1}{M}
\int d^4\theta\,\Big[
H_1^\dagger \,e^{\Vm} Q\,Y_U\,U^c\,
+H_2^\dagger\,e^\Vp  Q\,Y_D\,D^c\,
+H_2^\dagger\,e^\Vp L\,Y_E\,E^c\,+h.c.
\Big]\nonumber\\[8pt]
&+&
\frac{1}{M}
\int d^4\theta\,\,
\Big[
A(S,S^\dagger)\,D^{\alpha}\,\Big(\,B(S,S^\dagger)\,H_2 \,e^{-\Vm}\Big)\,
\,D_{\alpha}\,\Big(\Gamma(S,S^\dagger)\,e^\Vm\,H_1\,\Big)+h.c.\Big]
\label{dim5}
\eea

\medskip\noindent
where $T_{Q,L}$ carry four indices (2 for each up/down sector), and
\bea
T_Q\equiv T_Q(S),\,\,\,\, T_L\equiv T_L(S),\,\,\,\,\,
\lambda_H\equiv \lambda_H(S),\qquad
Y_F\equiv Y_F(S,S^\dagger),\,\,\,F:U,D,E\,
\eea
showing the spurion dependence of various couplings\footnote{More
  exactly, the notation in eq.(\ref{ll5}) stands for
$Q\,\, U^c\,T_Q\,Q\,D^c
\equiv (Q\,\, U^c)^T\,(i\sigma_2)\,T_Q\,Q\,D^c$.
Similarly,
$D^{\alpha}\,[\,B(S,S^\dagger)\,H_2 \,e^{-\Vm}]\,
\,D_{\alpha}\,[\Gamma(S,S^\dagger)\,e^{\Vm} H_1]
\equiv
D^{\alpha}\,[\,B(S,S^\dagger)\,H_2^T\,(i\sigma_2)\, e^{-\Vm}]\,
D_{\alpha}\,[\Gamma(S,S^\dagger)\,e^{\Vm} H_1 ]$.}.
In (\ref{dim5}), $M$ is a mass scale associated with the generation
of the dimension-five operators, for example the mass of some heavy
particles integrated out.
The operator $(H_1\,H_2)^2$ is easily generated by integrating out a
singlet\footnote{From a superpotential
 $\mu H_1 \,H_2+m\,\Sigma^2+ \lambda \,\Sigma \,H_1\,H_2$
integrating a singlet $\Sigma$ generates $\lambda_H (H_1\,H_2)^2$.}.
The remaining  operators in (\ref{ll5}) are shown to be generated
in Appendix~\ref{appendixB} (see also Appendix~\ref{appendixA}) by
integrating out two massive ($SU(2)$ doublets)
superfields of mass  of order $M$.\footnote{In Appendix~\ref{appendixA}
it is shown how  $H_2\,D^2\,H_1 \sim D^\alpha H_2\,D_\alpha \,H_1$
is generated by integrating a massive superfield without gauge interactions.
In the presence of gauge interactions one finds the last operator
in (\ref{ll5}) (Appendix\,\ref{appendixB}).}
Therefore these operators have a natural presence at low energies.
The spurion dependence associated to these operators is
the most general one can have.
Since we  assume a spontaneously broken effective
Lagrangian, consistency of the integrating out procedure
 implies the restriction
\be
M_s \ \ll \ M \ .
\ee
Also we have in general
\bea\label{definitions3}
A(S,S^\dagger)&=& \alpha_0+\alpha_1\,S+\alpha_2\,S^\dagger+
\alpha_3\,S\,S^\dagger\nonumber\\
B(S,S^\dagger)&=&\beta_0+\beta_1\,S+\beta_2\,S^\dagger+
\beta_3\,S\,S^\dagger\nonumber\\
\Gamma(S,S^\dagger)&=&
\gamma_0+\gamma_1\,S+\gamma_2\,S^\dagger+\gamma_3\,S\,S^\dagger
\eea

\noindent
 The Lagrangian in (\ref{L1}), (\ref{LMSSM}), (\ref{ll5})
 contains however redundant terms,  due to possible
field redefinitions which relate various operators as we shall see
shortly. Familiar
transformations are  holomorphic field redefinitions
\be
\Phi_i \ \rightarrow \ (1 - k_i \ S ) \ \Phi_i \ , \label{holom}
\ee
which are commonly used in MSSM in order to restrict the couplings of the
spurion $S$, and thus, the so-called soft-breaking terms. We shall
use this freedom later on. Less familiar  are the
following (super)field transformations\footnote{To avoid
a complicated index notation, the transformations
 in (\ref{tra}) are written in a matrix notation for the
 Higgs $SU(2)$ doublets,  thus the presence of
 $(i\sigma_2)$, although in the superpotential
 this is not shown explicitly.}
\medskip
\bea\label{tra}
H_1 \ \ra \ H_1' &=& H_1-\frac{1}{M}\,
\overline D^2\,\Big[\Delta_1\,H_2^\dagger\,
e^{\Vp}\,(i\,\sigma_2)\Big]^T
+\frac{1}{M} \,Q\,\rho_U\,U^c
\nonumber\\[10pt]
H_2 \ \ra \ H_2' &=& H_2
+
 \frac{1}{M}\, \overline
D^2\,\Big[\Delta_2\,H_1^\dagger\,e^\Vm\,(i\sigma_2)\Big]^T
+\frac{1}{M}\,Q\,\rho_D\,D^c+\frac{1}{M}\,L\,\rho_E\,E^c \eea
Here \bea
\rho_F=\rho_F(S);\,\,\,\,F:U,D,E,\,\,\qquad
\Delta_i=\Delta_i(S,S^\dagger)\qquad\,\,i=1,2 \label{ft}
\eea are
arbitrary functions of the spurion, i.e. their coefficients in the
Taylor expansion in $S$ are free parameters, which can be chosen to
eliminate redundant dimension-five operators, as we shall see shortly.
These coefficients should have values smaller than $M$ and the same
applies to the entries of the $\rho_F$, $F=U,D,E$ which are
$3\times 3$ matrices. We take
\bea\label{delta1and2}
\Delta_1(S,S^\dagger)&=&s_0+s_1\,S+s_2\,S^\dagger+s_3\,S\,S^\dagger
\nonumber\\
\Delta_2(S,S^\dagger)&=&s_0'+s_1'\,S+s_2'\,S^\dagger+s_3'\,S\,S^\dagger
\eea
Notice that in the R-parity violating MSSM, we would also have the
freedom to perform field transformations similar to  (\ref{tra}) on
quarks and leptons superfields. It is easy to see,
however, that all these new transformations, with the exception of
(\ref{tra}), violate R-parity
and cannot therefore be performed in the
R-parity conserving MSSM extension.
  Notice that field redefinitions (\ref{tra}), in addition of mixing
operators from ${\cal L}^{(4)}_{MSSM}$  and ${\cal L}^{(5)}$, also
 generate operators of  higher-order in $1/M$ (dimension-six), 
of the type
\medskip
\bea
\frac{1}{M^2} \, \int d^4 \theta \ D^2
\big[ H_2 \,e^{-\Vm} \Delta_1^{\dagger}\big]
\, e^{\Vm} \ {\bar D}^2
\big[ \Delta_1  \, e^{-\Vm}\, H_2^{\dagger} \big]
\eea

\medskip\noindent
plus a similar one for $H_1$. Since the effects of such operators are
further suppressed with respect to the dimension-five operators we are
considering, we shall neglect them in what follows.
One then finds that the  original Lagrangian transforms into:
\medskip
\bea\label{ll6}
\cL &=&\cL_K+
\int d^4\theta\,\,\Big[
  \cZ_1'\,H_1^\dagger \,e^\Vm\,H_1+
  \cZ_2'\,H_2^\dagger \,e^{\Vp}\,H_2\Big]
\nonumber\\[7pt]
&+&
\int d^2\theta\,\,
\Big[
- H_2\,Q\,\lambda'_U\,U^c
- \, Q\,\lambda'_D\,D^c\,H_1
- \, L\,\lambda'_E\,E^c\,H_1 +
\mu\,H_1\, H_2\Big]+h.c.
\nonumber\\[7pt]
&+&
\frac{1}{M}
\int d^2\theta\,\,
\Big[\,
Q\,U^c\,T_Q'\,Q\, D^c+Q\,U^c\,T_L'\,L\,E^c+
\lambda_H \,(H_1\,H_2)^2\,\Big]+h.c.
\nonumber\\[7pt]
&+&
\frac{1}{M}
\,\int d^4\theta\,\Big[
H_1^\dagger\,e^\Vm\,Q\,Y_U'\,U^c+
H_2^\dagger\,e^{\Vp} Q\,Y_D'\,D^c+
H_2^\dagger\,e^{\Vp} L\,Y_E'\,E^c+h.c.\Big]
+\Delta\cL\qquad
\eea
where\footnote{In a matrix notation, in
(\ref{deltao}) one replaces $H_2\ra H_2^T\,(i\sigma_2)$, and
similar  for the holomorphic part of (\ref{ll6}).}
\bea\label{deltao}
\Delta\cL&=&\frac{1}{M}
\int d^4\theta \,\Big[
-
\Delta_1^\dagger\,H_2 \,e^{-\Vm}  D^2 (\cZ_1 \ e^{\Vm} H_1)
-
\cZ_2\,H_2\,e^{-\Vm}\, D^2(\Delta_2^\dagger\,e^\Vm\,H_1)
+h.c.\Big]\nonumber\\[8pt]
&+&
\frac{1}{M} \int d^4\theta\,\, \Big[
A(S,S^\dagger)\,D^{\alpha}\,\big(\,B(S,S^\dagger)\,
H_2 \,e^{-\Vm}\big)\,\,D_{\alpha}\,
\big(\Gamma(S,S^\dagger)\,e^\Vm\,H_1\,\big)
+
h.c.\Big]\qquad
\label{redef}\eea

\bigskip
\noindent
Above we introduced the notation:
\medskip
\bea\label{lambdas}
\lambda'_F(S)=\lambda_F(S)+\frac{\mu(S)}{M}\,\rho_F(S),\qquad F:U,D,E
\eea
and
\bea\label{whys}
Y_U'(S,S^\dagger)&=&Y_U(S,S^\dagger)-4\,\Delta_2(S,S^\dagger)\,\lambda_U(S)
+\cZ_1(S,S^\dagger)\,\rho_U(S)\nonumber\\[8pt]
Y_D'(S,S^\dagger)&=&Y_D(S,S^\dagger)-4\,\Delta_1(S,S^\dagger)\,\lambda_D(S)
+\cZ_2(S,S^\dagger)\,\rho_D(S)\nonumber\\[8pt]
 Y_E'(S,S^\dagger)&=&
 Y_E(S,S^\dagger)-4\,\Delta_1(S,S^\dagger)\,\lambda_E(S)
+\cZ_2(S,S^\dagger)\,\rho_E(S)
\eea
and
\bea\label{ts}
 T'_Q(S)&=&T_Q(S) \ + \ \lambda_U(S)\, \otimes\, \rho_D(S) \ +
\rho_U(S)\, \otimes\, \lambda_D(S)\, \nonumber\\[8pt]
 T_L'(S)&=&T_L(S) \ + \ \lambda_U(S)\, \otimes \rho_E(S) \
 + \ \rho_U(S)\, \otimes  \lambda_E(S)
\eea
Finally
\bea\label{zs}
\cZ_1'(S,S^\dagger)&=&\cZ_1(S,S^\dagger)-\frac{1}{M}\,
\Big(4 \,\mu(S)\,
\Delta_2(S,S^\dagger)+h.c.\Big)
,\nonumber\\[8pt]
 \cZ_2'(S,S^\dagger)&=&
\cZ_2(S,S^\dagger)-\frac{1}{M}\, \Big(\, 4 \,\mu(S)\,
\Delta_1(S,S^\dagger)+h.c.\Big)
\eea

\medskip
\noindent
In eqs.(\ref{lambdas}), (\ref{whys}), (\ref{ts})
all quantities  except $M$ are  functions of the spurion field.
Next rescale the Higgs fields for canonical normalisation
of their kinetic terms
\bea\label{fields}
H_1&\ra& \frac{1}{\sqrt{a_0'}}\,
 \big[1-k_1\,S\big]\,H_1,\qquad
H_2\ra\frac{1}{\sqrt{b_0'}}\,
 \big[1-k_2\,S\big]\,H_2,\qquad k_1\equiv \frac{a_1'}{a_0'},\,\,\,
 k_2\equiv \frac{b_1'}{b_0'}
\eea
with
\bea\label{aandb}
a_0'\equiv\cZ_1'\Big\vert_{S,S^\dagger=0},\qquad
a_1'\equiv \cZ_1'\Big\vert_{S},\qquad
b_0'\equiv\cZ_2'\Big\vert_{S,S^\dagger=0},\qquad
b_1'\equiv\cZ_2'\Big\vert_{S}
\eea

\medskip\noindent
which can be immediately computed using
the definition of $\cZ_{1,2}'$, $\cZ_{1,2}$ and $\Delta_{1,2}$
and their spurion dependence given  above.
After the Higgs fields transformation we obtain

\bea\label{ll7}
\cL&=& \cL_K+\Delta\cL+
\int d^4\theta\,\,\Big[
  \Big(1- \frac{m_1^2}{M_s^2} \,S\,S^\dagger\Big) \,H_1^\dagger \,e^\Vm\,H_1
+\Big(1- \frac{m_2^2}{M_s^2} \,S\,S^\dagger\Big) \,H_2^\dagger \,e^{\Vp}\,H_2
\Big]\nonumber\\[8pt]
&+&
\int d^2\theta\,\,
\Big[
- H_2\,Q\,\lambda''_U\,U^c
- \, Q\,\lambda''_D\,D^c\,H_1
- \, L\,\lambda''_E\,E^c\,H_1 +
\mu'\,H_1\, H_2\Big]+h.c.
\nonumber\\[8pt]
&+&
\frac{1}{M}
\int d^2\theta\,\,
\Big[\,
Q\,U^c\,T_Q'\,Q\, D^c+Q\,U^c\,T_L'\,L\,E^c+
\lambda_H' \,(H_1\,H_2)^2\,\Big]+h.c.
\nonumber\\[8pt]
&+&
\frac{1}{M}
\,\int d^4\theta\,\Big[
H_1^\dagger\,e^\Vm\,Q\,Y_U''\,U^c+
H_2^\dagger\,e^{\Vp}\,Q\,Y_D''\,D^c+
H_2^\dagger\,e^{\Vp}\,L\,Y_E''\,E^c+
h.c.\Big]\qquad
\eea

\bigskip\noindent
Above we introduced the following
notation for the spurion dependent quantities:
\medskip
\bea\label{lambdas2}
\lambda_U''(S)&=&\frac{1}{\sqrt{b_0'}}\,
 \,(1-k_2\,S)\,\,\lambda_U'(S)=(1-b_1\,S)\,\lambda_U(S)+\cO(1/M),
\nonumber\\
\lambda_F''(S)&=& \frac{1}{\sqrt{a_0'}}\,
\,(1-k_1\,S)\,\,\lambda_F'(S)
=(1-a_1\,S)\,\lambda_{F}(S)+\cO(1/M),\qquad F \equiv D, E.
\nonumber\\
\mu'(S) &=&
\frac{1}{\sqrt{a_0'\,b_0'}}\,[1- (k_1+k_2)S]\,\,\mu(S)
=(1-(a_1+b_1)\,S)\,\mu(S)+\cO(1/M).
\eea

\medskip\noindent
Since $a_0', b_0'$ are $M$-dependent, see
(\ref{zs}), (\ref{aandb}),   the couplings
$\lambda_{U,D,E}''(S)$ and also $\mu'(S)$
have  acquired, already at the classical level,
a dependence on the scale $M$ of the higher dimensional
operators (threshold correction).
This is denoted above by $\cO(1/M)$ and  can
be easily computed using (\ref{zs}), (\ref{aandb}). Note that this
 $\cO(1/M)$  correction is relevant for the Lagrangian
(\ref{ll7}). Similar considerations apply to
$m_{1,2}$ entering in the first line in (\ref{ll7}) and
their exact expressions (not shown) in terms of initial parameters
can be computed in the same way.
Further
\medskip
\bea\label{yukawas}
\lambda_H'(S)\,=
 \Big(1-2 (a_1+b_1)\, S\Big)\,\,\lambda_H(S),&\,&
Y_U''(S,S^\dagger)\, =
\,\, (1-a_1^* \,S^\dagger\,)\,\,Y_U'(S,S^\dagger)
\nonumber\\[7pt]
Y_D''(S,S^\dagger) =
 (1-b_1^*\, S^\dagger\,)\,\,Y_D'(S,S^\dagger),\qquad &\,&
 Y_E''(S,S^\dagger) =
 (1-b_1^* \, S^\dagger\,)\,\,Y_E'(S,S^\dagger)\qquad\,\,\,
\eea

\medskip\noindent
where we ignored terms which bring $\cO(1/M^2)$
corrections to (\ref{ll7}).
Finally, $\Delta\cL$ in (\ref{ll7})  is that of
(\ref{deltao})  after applying to it
transformation (\ref{fields}). This  gives
\medskip
\bea\label{deltaL1}
\Delta\cL&
=&
-\frac{1}{M}
\int d^4\theta
\,\,t_0\,\,H_2\,e^{-\Vm} \,D^2 \,\Big[e^\Vm \, H_1\Big]
\nonumber\\[7pt]
&+&
\frac{M_s}{M}\,\Big[\,\,
4 \,\big[t_1+t_2+t_0(a_1+b_1)\big]\,h_2 \,
\cD_\mu\cD^\mu\,h_1
-
2 \, \big[t_1-t_2+t_0(b_1-a_1)\big]\,h_2\,D_1\,h_1\nonumber\\[7pt]
&+&
2\sqrt 2 \,(t_1+b_1\,t_0)\,h_2\,\lambda_1\,\psi_{h_1}
-
2\sqrt 2 \,(t_2+a_1\,t_0)\,\psi_{h_2}\,\lambda_1\,h_1
- 4\,t_3\,F_{h_2}\,F_{h_1}
\Big]\nonumber\\[7pt]
&+&
\frac{M_s^2}{M}\,\Big[
-4 \,(t_4-b_1\,t_3)\, h_2\,F_{h_1}
-4 \,(t_5-a_1\,t_3)\,F_{h_2}\,h_1+
2\,t_6\,\psi_{h_2}\psi_{h_1}
\Big]\nonumber\\[7pt]
&+&
\frac{M_s^3}{M}\,\Big[-4\,(t_7-a_1\,t_4-b_1\,t_5+a_1 \,b_1 \,t_3)
\,\,h_2h_1\Big]+h.c.
\eea

\bigskip\noindent
where the hermitian conjugation h.c. applies to all terms above and where we
ignored $\cO(1/M^2)$ corrections. Also $D_1$ and $\lambda_1$
are components of the vector superfield $V_1$ and we also used the
component notation $H_i=(h_i,\psi_{h_i},F_{h_i})$.
In $\Delta\cL$ we replaced  $k_{1}, (k_2)$ by $a_1$, ($b_1$) respectively,
which is correct in the approximation of ignoring $1/M^2$ terms in the
Lagrangian. The coefficients $t_i$ are given by

\bea
t_0&=&\alpha_0\beta_0\gamma_0
+ s_0^*+ s_0^{' *},
\hspace{2.6cm}
t_4= d_4- 
\,s_3^*-a_1^*\,s_2^*- b_2\,s_0^{' *} -b_1\,s_1^{' *},
\nonumber\\[7pt]
t_1&=&d_1-
s_2^* - b_1\,s_0^{' *},
\hspace{2.9cm}
t_5=d_5-a_2\,s_0^*-a_1\,s_1^*-
s_3^{' *}- b_1^*\,s_2^{' *},
\nonumber\\[7pt]
t_2&=&d_2-a_1\,s_0^*-
s_2^{' *},
\hspace{2.8cm}
t_6=d_6,
\nonumber\\[7pt]
t_3&=&d_3-
s_1^*- a_1^*\,s_0^*- 
s_1^{' *}- b_1^*\,s_0^{' *},
\hspace{0.7cm}
t_7=d_7-a_2\,s_2^*-a_1\,s_3^*-b_1\,s_3^{' *}-b_2 \,s_2^{' *}
\eea

\medskip\noindent
and where $d_i$ are combinations of input parameters
$\alpha_i, \beta_i,\gamma_i$ of eq.(\ref{definitions3})
\bea\label{ds}
d_1&\equiv& - \beta_1\, \alpha_0\,\gamma_0\,
-\,\alpha_1\,\beta_0\,\gamma_0/2,
\hspace{2cm}
d_4\equiv  -\beta_3\,\alpha_0\,\gamma_0-\beta_1\,\alpha_2\,\gamma_0
-\alpha_0\beta_1\gamma_2
\nonumber\\[7pt]
d_2&\equiv&  -\gamma_1\,\beta_0\,\alpha_0
-\,\alpha_1\,\beta_0\,\gamma_0/2,
\hspace{2.cm}
d_5 \equiv -\gamma_3\,\beta_0\,\alpha_0-\gamma_1\,\alpha_2\,\beta_0
-\alpha_0\beta_2\gamma_1,
\nonumber\\[7pt]
d_3 &\equiv &-\alpha_2\,\beta_0\,\gamma_0
-\alpha_0\beta_2\gamma_0-\alpha_0\beta_0\gamma_2,
\hspace{0.9cm}
d_6 \equiv \alpha_3\,\gamma_0\,\beta_0
+\alpha_1\beta_2\gamma_0+\alpha_1\beta_0\gamma_2
\nonumber\\[7pt]
&&\hspace{5.9cm}
d_7\equiv -\gamma_3\,\beta_1\,\alpha_0-\gamma_1\,\beta_3\,\alpha_0-
\gamma_1\,\beta_1\,\alpha_2. \qquad
\eea

\medskip\noindent
A suitable choice of coefficients  $s_0, s_0', s_2', s_2$
entering in transformation (\ref{tra}) allows us to set
\bea\label{constraints}
t_i=0,\qquad i=0,1,2, 3.
\eea
This ensures that the non-standard terms in 
the first, second and third  lines of
$\Delta\cL$ above are not present. The remaining terms
proportional to $M_s^2$ and $M_s^3$
bring a renormalisation of the
soft terms only,
which are present anyway in the Lagrangian
of (\ref{ll7}), thus can be ignored
(recall that the auxiliary fields
can be replaced onshell by their lowest order (MSSM) values).
Finally, the term $t_6\, \psi_{h_2}\psi_{h_1}$
brings a renormalisation of the supersymmetric $\mu'$ term ($\mu'
H_1 H_2$) of (\ref{ll7}), induced by soft supersymmetry breaking,
and is invariant under the general field transformations (\ref{tra}).
In principle one could set additional coefficients of the last two
lines in $\Delta\cL$
to vanish by a suitable choice of remaining $s_{1,3}, s_{1,3}'$; 
we choose not to do so and instead  save these remaining coefficients
 for additional conditions that can be used to simplify the
(couplings or the spurion  dependence of our)  Lagrangian 
even further.

We then obtain the minimal set of
dimension-five operators beyond the MSSM Lagrangian
\smallskip
\bea
\label{FL}
\cL&=&\cL_K+
\int d^4\theta\,\,\Big[\Big(
1- \frac{m_1^2}{M_s^2} S^{\dagger} S\Big) \,H_1^\dagger \,e^\Vm\,H_1
+ \Big(1- \frac{m_2^2}{M_s^2} S^{\dagger} S\Big)
\,H_2^\dagger \,e^{\Vp}\,H_2 \Big]\nonumber\\[8pt]
&+&
\int d^2\theta\,\,
\Big[
- H_2\, Q\,\lambda_U'' (S)\,U^c
- Q\,\lambda_D'' (S) \,D^c\,H_1
- L\,\lambda_E'' (S) \,E^c\,H_1
+ \mu'' (S) \,H_1\, H_2\Big]+h.c.
\nonumber\\[8pt]
&+&
\frac{1}{M}
\int d^2\theta
\,\,
\Big[\,
Q\,U^c\,T_Q'(S) \,Q\, D^c+Q\,U^c\,T_L' (S) \,L\,E^c+
\lambda_H' (S) \,(H_1\,H_2)^2\,\Big]+h.c.
\nonumber\\[8pt]
&+&\!\!\!\!
\frac{1}{M}\!
\,\int\! d^4\theta\,\Big[
H_1^\dagger\,e^\Vm  Q\,Y_U''(S,S^{\dagger})\,U^c\!
+
H_2^\dagger\, e^\Vp Q\,Y_D''(S,S^{\dagger})\,D^c\!
+
H_2^\dagger\, e^\Vp L\,Y_E''(S,S^{\dagger})\,E^c\!+h.c.\Big]
\nonumber\\
\label{final}
\eea

\noindent
where  $\cL_K$ stands for gauge kinetic terms and
 for kinetic terms of MSSM fields other than
$H_{1,2}$, together with their spurion dependence;
$\mu''$ now includes the renormalisation due to $t_6$ (not shown).
This Lagrangian gives the  irreducible
set of dimension-five R-parity conserving operators that can be present
beyond the MSSM and is one of the main results of this work. As explained
above, there is still some remaining freedom of the
field redefinitions that will be used in the next 
section. The couplings entering above
are given in eqs.(\ref{lambdas}), (\ref{whys})
(\ref{ts}), (\ref{lambdas2}), (\ref{yukawas})
in terms of those in the original Lagrangian.
The couplings $\lambda''_{U,D,E}(S)$ acquired a threshold correction
$\cO(1/M)$, which can be obtained from (\ref{lambdas2}).
The dimension-five operator that was present in the last line
of (\ref{ll5})
was completely ``gauged away'' in the new fields basis,
up to effects which renormalised the soft terms (unknown anyway)
or the supersymmetric $\mu$ term. Since the physics
should be independent of the fields basis,
in this new basis it is manifest that the last operator in
 (\ref{ll5}) cannot affect the relations among physical masses
of the Higgs sector. We discuss this in detail in Section~\ref{ss2}.

\section{Phenomenology of the new couplings of the MSSM$_5$.}
\label{ss1p}

The Lagrangian in (\ref{FL}) has couplings which can
have dramatic implications if the scale $M$ is not
high enough, in particular due to FCNC effects.
Indeed, if $T'_{Q,L}$ and $Y_F''$, $F:U,D,E$, of  (\ref{FL})
have arbitrary family dependent 
couplings, one expects stringent limits from FCNC bounds 
\cite{Gabbiani:1996hi}.  It is possible however,
 under some  mild  assumptions for the original
$\cL$ of (\ref{L1}) with (\ref{LMSSM}), (\ref{ll5}),
that  some of the couplings in (\ref{FL}) can
be also removed. For example assume that in the 
original Lagrangian (\ref{ll5})
all flavor matrices are proportional to the ordinary Yukawa couplings
and similar
for\footnote{The eqs in (\ref{ansatz0}) are also motivated by
the discussion in Appendix~\ref{appendixB}, eq.(\ref{B7}) where a
similar structure of  $T_{Q,L}$ and $\rho_F$ is generated by integrating
out massive $SU(2)$ superfields doublets.}
$\rho_F$ of (\ref{tra}), (\ref{ft}): 
\medskip
\bea\label{ansatz0}
T_Q(S)& = & c_Q(S)\,\,\lambda_U(0)\otimes \,\lambda_D(0)\nonumber\\
T_L(S)& = & c_L(S)\,\,\lambda_U(0)\otimes \,\lambda_E(0)\nonumber\\
\rho_F(S)& = & c_F(S)\,\,\lambda_F(0),\,\,\,\,\,\,\,\, F: U,D,E
\eea
and, as usual
\bea\label{lambdas3}
\lambda_F(S)&=&\lambda_F(0)\,(1+A_F \,S),\,\,\,\,\, F: U,D,E.
\eea

\medskip\noindent
Above $c_{Q,L}(S)$ are some arbitrary
input functions of $S$; $\lambda_F(S)$
with $F: U,D,E$ are  $3\times 3$ matrices,
while $A_F$ are trilinear couplings.
In the following
$c_F(S)\equiv c_0^F+S\,\,c^F_1$, $F=U,D,E$ are regarded as free
parameters which can be adjusted, together with the remaining\footnote{
see (\ref{tra}), (\ref{delta1and2}) and (\ref{constraints}).}
$s_{1,3}$, $s_{1,3}'$, to remove some of the couplings in (\ref{FL}).
Indeed, if
\bea
c_U(S)=- c_L(S)-c_E(S), \quad
c_D(S)=- c_Q(S)+c_L(S)+c_E(S)
\eea
while  $c_E(S)$ remains arbitrary, one obtains
\bea \label{TQL}
T_Q'(S)=0,\quad  T_L'(S)=0
\eea

\medskip\noindent
We can therefore remove the associated couplings in (\ref{FL}),
the first two terms in the third line of (\ref{FL}).
Finally, let us assume that in (\ref{ll5}) we also have
\bea\label{ansatz1}
Y_F(S,S^\dagger)=\,f_F(S,S^\dagger)\,\lambda_F(0),\qquad F:U,D,E
\eea
where $f_F$ are spurion-dependent, family-independent functions of arbitrary
coefficients:
\bea
f_F(S,S^\dagger)=f_0^F+ S\,f_1^F+ S^\dagger\,f_2^F
+S\,S^\dagger\,f_3^F
\eea
Using (\ref{yukawas}),
we find that the couplings in (\ref{FL}) are
\medskip
\bea
Y_F^{''}(S,S^\dagger) =
\lambda_F(0)\,\Big[ x^F_0 + x^F_1 \,\,S
+ x^F_2\,\,S^\dagger\,
+ x^F_3  \,\,S\,S^\dagger\Big],\quad
F=U,D,E
\eea

\medskip\noindent
One finds
\bea
x^U_0&= &f^U_{0}- 4 s_0' +  \,c_0^U
\nonumber\\
x^U_1&= &f^U_{1}-4\,s_1'+ \,c^U_1+a_1\,c_0^U
\nonumber\\
x^U_2&=&f^U_2-4\,s_2'+a_1^*\,c^U_0- 
a_1^*\, x_0^U
\nonumber\\
x^U_3&=&f^U_3-4\,s_3'+a_1^*\,c^U_1+a_2\,c^U_0 - 
a_1^* \,\,x_1^U
\eea

\medskip\noindent
Similar equations exist for $D$ fields, obtained from those above
with replacements $U\ra D$, $s_i'\ra s_i$ and $a_i\ra b_i$.
Also for $E$ fields the replacements are
$U\ra E$, $s_i'\ra s_i$ and $a_i\ra b_i$.

Let us examine  if the form of $Y''_F(S,S^\dagger)$
can  be simplified using the free parameters that we are left
with: these  are $s_{1,3}, s_{1,3}'$  from general transformations
$\Delta_{1,2}$ and $c_E(S)=c^E_0+S\,c^E_1$ thus a total of 6 free
parameters. We can use $s_{1,3}'$  ($s_{1,3}$)
to eliminate $S$ and $S\,S^\dagger$ parts of
$Y_U''$  \,\, ($Y_D''$), respectively.
Using $c_0^E$ and $c_1^E$
we can also eliminate the  $S$ and $S\,S^\dagger$ of $Y_E''$.
In conclusion we used the remaining
 6 free parameters to bring $Y''_F$ to the form
\medskip
\bea\label{newY}
Y''_F(S^\dagger)\equiv Y_F''(0,S^\dagger)
=\lambda_F(0)\,(x_0^F+ x_2^F\,\,S^\dagger),\qquad
F:U,D,E
\eea

\medskip\noindent
The coefficients $x_{0,2}^F$ depend on the arbitrary
(input) coefficients  $f_i^F$, $i=0,1,2,3$,  $a_i$, $b_i$, $c_i$
of the original Lagrangian (\ref{L1}), (\ref{LMSSM}), 
(\ref{ll5}). Other simplifications can occur
if we ignore the couplings $Y$ of the first two families.
With these considerations,
 the Lagrangian in (\ref{FL}) becomes
\medskip
\bea\label{lastL}
\cL&=&\cL_K+
\int d^4\theta\,\,\Big[
\Big(1-  \frac{m_1^2}{M_s^2} S^{\dagger} S\Big) 
\,H_1^\dagger \,e^\Vm\,H_1
+
\Big(1- \frac{m_2^2}{M_s^2} S^{\dagger} S\Big) 
\,H_2^\dagger \,e^{\Vp}\,H_2 \Big]
\nonumber\\[7pt]
&+&
\int d^2\theta\,\,
\Big[
- H_2\, Q\,\lambda_U'' (S)\,U^c -
Q\,\lambda_D'' (S) \,D^c\,H_1-
L\,\lambda_E'' (S) \,E^c\,H_1+\mu'' (S) \,H_1\, H_2\Big]+h.c.
\nonumber\\[7pt]
&+&\!\!\!\!
\frac{1}{M}\!
\,\int\! d^4\theta\,\Big[
H_1^\dagger\,e^\Vm\,Q\,Y_U''(S^{\dagger})\,U^c
+
H_2^\dagger\,e^{\Vp} Q\,Y_D''(S^{\dagger})\,D^c
+
H_2^\dagger\,e^{\Vp}  L\,Y_E''(S^{\dagger})\,E^c+h.c.\Big]\qquad
\nonumber\\[7pt]
&+&\frac{1}{M}\int
d^2\theta\,\,\lambda_H'(S)\,(H_1\,H_2)^2+h.c.
\label{final2}
\eea

\bigskip\noindent
with  couplings (\ref{newY}), (\ref{lambdas2})\footnote{
$\lambda_F''(S)$ acquired a threshold correction in $M$:
$\lambda''_U(0)=\lambda_U(0)\,
\big[1+{1}/{M}\,\big(\mu(0)\,c_U(0)+2\,(\mu(0)\,s_0+\mu^*(0)\,s_0^*
)\big)\big]
$
with similar relations for  $D$, $E$
 obtained by
$s_0\ra s_0'$ and  $U\ra D$, ($U\ra E$).
In terms of original parameters, 
$
s_0=-[-4 \alpha_0^*\beta_0^*\gamma_0^*\,b_1-4
\,d_3^*+
(f_1^U+f_1^D+c_1^U+c_1^D+a_1\,c_0^U+b_1\,c_0^D)]/
4\,(a_1-b_1)$
with $d_3$ as in (\ref{ds}); for the $D, E$ sectors we use
$s_0'=-\alpha_0^*\beta_0^*\gamma_0^*-s_0$.
Similar relations exist for
non-supersymmetric counterparts, see (\ref{lambdas2}),
(\ref{yukawas}).}. This defines our MSSM extension
with D=5 operators (MSSM$_5$).

A detailed analysis of all couplings
generated by (\ref{lastL}) or by (\ref{FL}) and their phenomenological
implications is beyond the scope of this paper.
For related studies see also the 
analysis in \cite{Buchmuller,Piriz:1997id,Pospelov:2005ks}.
Let us present however all the new couplings generated
using component fields and
we begin with the couplings proportional to $M_s$.
Part of these are coming
from the terms in the second-last line of (\ref{lastL}).
These  include non-analytic  Yukawa couplings \cite{M0}
\medskip
\bea\label{cset3}
&& \frac{M_s}{M}\,x_2^U\,(\lambda^U_0)_{ij}\,\,
(h_1^\dagger\,q_{L\,i})\,\,u_{R\,j}^c+h.c.\nonumber\\
&&  \frac{M_s}{M}\,x_2^D\,(\lambda^D_0 )_{ij}\,\,
(h_2^\dagger\,q_{L\,i})\,\,d_{R\,j}^c+h.c.\nonumber\\
&& \frac{M_s}{M}\,x_2^E\,(\lambda^E_0 )_{ij}\,\,
 (h_2^\dagger\,l_{L\,i})\,\,e_{R\,j}^c+h.c.,\qquad \lambda^F_0\equiv
\lambda_F(0),\,\,\, F:U,D,E.
\eea

\medskip\noindent
These  couplings
are not soft in the sense of \cite{Girardello:1981wz},
but ``hard'' supersymmetry breaking terms (for
 ``non-standard'' and ``hard'' supersymmetry
 breaking terms see \cite{M0,M1}); they are  less suppressed
than those listed in \cite{M0} where they were generated
at order $M_s^2/M^2$.
 Such  couplings can bring about a $\tan\beta$
enhancement of a prediction for a physical observable,
such as the bottom quark mass relative to bottom quark
Yukawa coupling \cite{Haber, Carena:2001bg}.
This effect  is also present  in the electroweak scale effective
  Lagrangian of the MSSM alone,
after integrating out massive squarks at one-loop level,
with a result for bottom quark mass
\cite{Haber,Carena:2001bg,Pierce:1996zz,Hall:1993gn,Carena:1994bv}
\bea
m_b=\frac{v\cos\beta}{\sqrt
  2}\,\Big( \lambda_b +{\delta}{\lambda_b}+
{\Delta}{\lambda_b}\tan\beta\Big)
\eea

\medskip\noindent
where $\lambda_b$ is the ordinary bottom quark
Yukawa coupling, $\delta\lambda_b$ its one loop correction
 and $\Delta\lambda_b$ is a ``wrong''-higgs
bottom quark  Yukawa coupling, generated by integrating out massive
squarks. In our case, $\Delta\lambda_b$ receives an additional
contribution from the  second line in (\ref{cset3}). The size of
this extra contribution due to  higher dimensional operators, can be
comparable and even substantially larger than the one generated
in the MSSM at one-loop level (for a suitable value for
$x_2^D\,M_s/M$ - recall that $x_2^D$ is not fixed). Such
contributions  can bring a $\tan\beta$ enhanced correction of the
Higgs decay rate to bottom quark pairs. Similar considerations apply
to the $U$ and $E$ sectors.

Other similar  couplings derived from (\ref{lastL})
and proportional to $M_s$ are
\medskip
\bea\label{cset4}
&&
\frac{M_s}{M}\,x_2^U\,(\lambda_0^{U\dagger} \lambda_0^U)_{ij}
\,\, (h_1^\dagger\,h_2^\dagger)\,\,\tilde u_{R\,i}\,\tilde u_{R\,j}^*
+h.c.\nonumber\\
&&
\frac{M_s}{M}\,x_2^U\,(\lambda_0^U\, \lambda_0^{U\dagger})_{ij}
\,\, (h_1^\dagger\,\tilde q_{L\,i})\,\,(h_2^\dagger 
\,\tilde q_{L\, j}^\dagger)
+h.c.
\eea

\medskip\noindent
where we used that
$\lambda_0^{F''}$ and $\lambda_0^{F}$ are equal
up to $\cO(1/M)$ corrections,
see (\ref{lambdas}), (\ref{lambdas2}).
The above terms are strongly suppressed due to the square
of the Yukawa coupling, in addition to $M_s/M\ll 1$,
so their effects are expected to be small, except for the third
generation.
Their counterparts in the down ($D$) sector are
\medskip
\bea\label{cset5}
&&
\frac{M_s}{M}\,x_2^D\,(\lambda_0^{D\dagger}\, \lambda_0^{D})_{ij}
\,\, (h_2^\dagger\,h_1^\dagger)\,\,\tilde d_{R\,i}\,\tilde d_{R\,j}^*
+h.c.\nonumber\\
&&
\frac{M_s}{M}\,x_2^D\,(\lambda_0^D\, \lambda_0^{D\dagger})_{ij}
\,\, (h_2^\dagger\,\tilde q_{L\,i})\,\,(h_1^\dagger
 \,\tilde q_{L\, j}^\dagger)
+h.c.
\eea

\medskip\noindent
In the lepton sector similar couplings
are present, obtained  from eq.(\ref{cset5})
with $Q\ra L$, $D\ra E$.
All the quartic couplings listed above are renormalisable, 
but naively they would seem to
break supersymmetry in a hard way if inserted into
loops with a cutoff larger than $M$. This is of course just an
artifact of using a cutoff larger than the energy scale of heavy
states that we integrated out.

It is  interesting to note that there is no
``wrong-Higgs''-gaugino-higgsino coupling generated \cite{M0}, even
though the original Lagrangian in eq.(\ref{ll5})  included it, see
eq.(\ref{deltaL1}) where
\medskip
\bea
\frac{M_s}{M}\,\,\big( \psi_{h_2}\,\lambda_1\,\,h_1+
h_2\,\lambda_1\,\,\psi_{h_1}\big)+h.c.
\eea

\medskip\noindent
was present. Such a coupling can  be generated
at one loop level, for a discussion see \cite{Haber}. This coupling was
removed in our case by a suitable transformation for the
Higgs fields (\ref{tra}). This  shows that not all
``wrong''-higgs couplings are actually independent
(this  may also apply when such couplings are generated at the loop
level). 
 
Note that in the MSSM$_5$  defined by eq.(\ref{lastL}),
couplings proportional to $M_s$ involving ``wrong''-higgs A-terms
are not present, given our ansatz (\ref{ansatz0}), (\ref{ansatz1}) 
leading to (\ref{newY}). If
this ansatz is not imposed on the third generation,
then  one could have such terms from  (\ref{FL})
\medskip
\bea
\frac{M_s^2}{M}\,\Big[
y_{u,3}\,
 h_1^\dagger\,\tilde q_{L,3}\,\,\tilde u_{R,3}^*
+
\,y_{d,3}\, 
\,h_2^\dagger\,\tilde q_{L,3}\,\,\tilde d_{R,3}^* 
+
 y_{e,3}\,\, 
h_2^\dagger\,\tilde l_{L,3}\,\,\tilde e_{R,3}^* \Big]
\eea

\medskip\noindent
where $y_{f,3}$, $f=u,d,e$ are the coefficients of
 component $S\,S^\dagger$ of $Y''(S,S^\dagger)$ of third generation.

There are also new, and perhaps most important,
 supersymmetric couplings generated, that
affect the amplitude of processes like
quark + quark $\ra$ squark + squark, or involving (s)leptons too.
These are
\medskip
\bea\label{qqsqsq} &&
\frac{1}{M}\,x_0^U\,(\lambda_0^D)_{ij}\,(\lambda_0^U)_{kl}
\,\,\tilde q_{L\,i}\,\tilde d_{R\,j}^*\,\,
q_{L\,k}\,u_{R\,l}^c+ h.c.
\nonumber\\
 && \frac{1}{M}\,x_0^D\,(\lambda_0^U)_{ij}\,(\lambda_0^D)_{kl}
 \,\,\tilde q_{L\,i}\,\tilde u_{R\,j}^*\,\, q_{L\,k}\,d_{R\,l}^c+h.c.
 \nonumber\\
&&\frac{1}{M}\,x_0^U\,(\lambda_0^E)_{ij}\,(\lambda_0^U)_{kl}
\,\,\tilde l_{L\,i}\,\tilde e_{R\,j}^*\,\, q_{L\,k}\,u_{R\,l}^c
+(L\leftrightarrow Q, E\leftrightarrow U)+h.c.
\label{squarkproduction}
\eea

\medskip\noindent
These couplings
can be important particularly for the third generation.
The largest effect would be for squarks pair production
from a pair of quarks; the process could
be comparable to the MSSM tree level
 contribution to the amplitude of the same process
\cite{Dawson:1983fw}. Indeed, let us focus on the $q {\bar q}
\rightarrow {\tilde q} {\tilde q^*}$ in MSSM generated by a tree-level
gluon exchange. The MSSM amplitude behaves as
\medskip
\begin{equation}
A_{q {\bar q} \rightarrow g \rightarrow {\tilde q} {\tilde q^*}} \sim
{\frac{g_3^2}{\sqrt{s}}} \ ,
\end{equation}

\medskip\noindent
where $s$ is the Mandelstam variable. On the other hand, the
operators (\ref{squarkproduction}) generate a contact term
contributing
\medskip
\begin{equation}
A_{q {\bar q} \rightarrow {\tilde q} {\tilde q^*}}^{MSSM_5} \sim
\frac{\lambda_0^U \lambda_0^D}{M} \ .
\end{equation}

\medskip\noindent
The dimension-five operator for the third generation has therefore a
comparable contribution to the MSSM diagrams for energies $E \ge
g_3^2 M$, which can be in the TeV range. In MSSM there are other
diagrams contributing to this process, in particular Higgs exchange.
It can be checked however that at energies above the CP-even
Higgs masses, the MSSM amplitude decreases in energy whereas the
contact term coming from the dimension-five operators gives
a constant contribution which is sizeable
for high energy. Of course, at energies above $M$ we should replace the
contact term by the corresponding tree-level diagram with exchange of
massive $SU(2)$ doublets (or whatever other physics generates
 this effective operator).

 Note that couplings similar to (\ref{qqsqsq}) could
also be generated by the term $\int d^2\theta\, (Q U)\,T_Q (Q D)$ of
(\ref{FL}). This term is not present in MSSM$_5$ of
(\ref{lastL})
 due to our FCNC ansatz
(\ref{ansatz0}), (\ref{TQL}); however, for the third generation
this constraint of the ansatz can be relaxed. Therefore the above process
of squark production can have an even larger amplitude,
 from contributions in the third line of (\ref{FL}).

The Lagrangian (\ref{lastL}) also contains other 
 (supersymmetric) couplings involving
gauge interactions which can be important for
 phenomenology. They arise from any dimension-five D-term in 
 (\ref{lastL}) giving

\bea\label{gg1}
\cL\! &\supset&\!\! \frac{(\lambda_0^U)_{ij} x_0^U}{M}\,
\Big[
-h_1^\dagger \,\cD_\mu\cD^\mu \,(\tilde q_{L\,i} \,\tilde u^*_{R\, j})
-
\frac{1}{\sqrt{2}}\,
h_1^\dagger \lambda_1\,\big(\,\tilde q_{L\,i}\,\,u_{R\,j}^c
+q_{L\, i} \,\,\tilde u_{R\,j}^*\big)
-\frac{1}{\sqrt 2}\,
\overline \psi_{h_1}\,\overline\lambda_1\,\tilde q_{L\,i}\,\,\tilde
u_{R\,j}^*\nonumber\\[4pt]
&+&
\frac{1}{2}\,\,
h_1^\dagger \,D_1\,\tilde q_{L\,i}\, \tilde u_{R\,j}^*
+
i\overline \psi_{h_1}\,\overline\sigma^\mu\,\cD_\mu \,\big(\tilde
q_{L\,i}\,\,u_{R\,j}^c+ q_{L\,i} \,\,\tilde u_{R\,j}^*\big)
\Big]\nonumber\\[7pt]
&+&(U\rightarrow D, \,H_1\rightarrow H_2,
 \,V_1\rightarrow V_2)+
(Q\rightarrow L, \,H_1\rightarrow H_2, 
\,V_1\rightarrow V_2, U\rightarrow E)+h.c.
\eea

\bigskip\noindent
where $D_1$, $\lambda_1$
are the auxiliary  and gaugino components of $V_1$ vector superfield,
and

\bea
D_1&\equiv& -\frac{g_2^2}{2}\,
\Big[\,h_1^\dagger\,\vec\sigma\,h_1
+
h_2^\dagger\,\vec\sigma\,h_2
+
\tilde q_{L\,i}^\dagger  \vec\sigma\tilde q_{L\,i}
+
\tilde l_{L\,i}^\dagger  \vec\sigma\tilde l_{L\,i}
\Big]\nonumber\\[5pt]
&+&
\frac{g_1^2}{2}
\Big[-h_!^\dagger h_1 +h_2^\dagger h_2+\frac{1}{3}\,\tilde
q_{L\,i}^\dagger
\tilde q_{L\,i}-\frac{4}{3}\tilde u_{R\,i}\tilde u_{R\,i}^*
+\frac{2}{3}\,\tilde d_{R\,i}\,\tilde d_{R\,i}^*
-\tilde l_{L\,i}^\dagger\,\tilde l_{L\,i}+2 \,\tilde e_{R\,i}\,\tilde
e_{R\,i}^*
\Big]
\eea

\bigskip\noindent
Here $\cD_\mu$ is the covariant derivative, $\cD_\mu=\partial_\mu
+i/2\,V_{1,\mu}$, where $V_{1,\mu}$ is the gauge field of the vector
superfield 
$\Vm\equiv \,g_2\, V_W^i\, \sigma^i -g_1 \, V_Y$, introduced
in eq.(\ref{LMSSM}).
 Couplings similar to those above are generated by
the substitutions shown in (\ref{gg1}).
Of the couplings above, phenomenologically relevant
could be those involving 2 particles and 2 sparticles, such as
higgs-quark-squark-gaugino, or gauge-quark-higgsino-squark
arising from (\ref{gg1}).  Also notice the presence in this eq
of the first term with  a ``wrong-higgs''-squark-squark 
derivative coupling.

Yukawa interactions also generate supersymmetric 
couplings of structure similar to some of those in (\ref{gg1}),
involving 4 squarks and a higgs or 2
squarks and 3 higgses, or 2 squarks, 2 sleptons plus a higgs. However,
these arise at order $\lambda_F^3$, where $\lambda_F$, $F:U,D,E$
are Yukawa couplings entering (\ref{lastL}).
Therefore they are suppressed both by the scale $M$ and,
relative to the above gauge counterparts, also by an extra Yukawa 
coupling (this is due to the presence of an extra Yukawa
 coupling in the third line
of (\ref{lastL}) relative to ordinary D-terms.
The strength of these interactions
is  also sub-leading to other Yukawa interactions listed so far
(which also  involved fewer (s)particles).

Finally, supersymmetric couplings with 3 higgses  
and 2 squarks or 2 sleptons arise from $(H_1 H_2)^2$ of (\ref{lastL}),
(suppressed by two Yukawa couplings and by the scale  $M$);
also generated are
potentially larger couplings of 2 higgses and 2 higgsinos, being
suppressed only by $\lambda_H(0)$ and by the scale $M$.
There are also non-supersymmetric  couplings with 4 higgs fields, 
whose effects are discussed in Section~\ref{ss2}.
This concludes our discussion of all the new couplings generated
by dimension-five operators in the MSSM$_5$.

\section{The MSSM Higgs sector
with dimension-five operators.}\label{ss2}

In the following we  restrict the analysis to the MSSM
Higgs sector extended by D=5 operators and analyse their
implications.
In this  sector there are in general two dimension-five operators
that can be present and affect the Higgs fields masses,
shown in eq.(\ref{lls}) below.
According to our previous discussion
the last operator in (\ref{lls}) is redundant and can be
``gauged away''.   However, in this section
we choose to keep it, in order to show explicitly
that it does not bring new physics of its own\footnote{
In the exact susy case, if set onshell  this operator brings only
wavefunction renormalisation
 (Appendix~\ref{appendixB})}.
The relevant part of  MSSM Higgs Lagrangian
with  D=5 operators is
\medskip
\bea\label{lls}
\cL_1&=&\int d^4 \theta \,\,\Big[
\cZ_1(S, S^\dagger)
\,\, H_1^\dagger \,e^{\Vm}\,H_1
+
\,\, \cZ_2(S,S^\dagger)
\,\, H_2^\dagger \,e^{\Vp}\,H_2 \Big]\\[8pt]
&+&
\int d^2\theta \,\,\Big[\,\tilde\mu\,\, (1+c_1\,S)\,\,H_1\,H_2
+ \,\frac{c_3}{M}
\,\,\,(1+c_2\, S)\,(H_1\,H_2)^2 \Big]+h.c.\nonumber\\[8pt]
&+&\!\!\!\!
\frac{1}{M}\!
\int\! d^4\theta\,\,\Big\{
A(S,S^\dagger)\,D^\alpha\,\Big[B(S,S^\dagger)\,H_2\,
e^{-\Vm}\,\Big]
D_\alpha\,\Big[\Gamma(S,S^\dagger)\,
e^{\Vm}\,H_1\,\Big]+h.c.\Big\}\nonumber
\eea

\bigskip\noindent
Additional spurion dependence arises from
the dimension-five  operators considered.
For the definitions of  $A(S,S^\dagger)$,
$B(S,S^\dagger)$, $\Gamma(S,S^\dagger)$ see eq.(\ref{definitions3}).
After some  calculations,
elimination of  the auxiliary fields and
a re-scaling of the scalar fields,
the scalar part of $\cL_1$ in (\ref{lls}) becomes:
\medskip
\bea\label{finalscalar}
\cL_{1,scalar}&=&
-\frac{1}{8}\,(g_1^2+g_2^2)\,
\big(\vert h_1\vert^2-\vert h_2\vert^2\big)^2
+\frac{M_s}{M}\,(g_1^2+g_2^2)\,\,\big(\vert h_1\vert^2-\vert
  h_2\vert^2\big)\,
\big(\delta_1\,h_1\,h_2\,+h.c.\big)\nonumber\\[8pt]
&+&
\frac{2\,c_3}{M}\,\,\big(\vert h_1\vert^2+\vert h_2\vert^2\big)
\big(\tilde\mu^*\,h_1\,h_2\,+h.c.\big)
-\frac{M_s}{M}\,\,c_3\, \big(\delta_2\, (h_1\,h_2)^2 +h.c.\big)
\\[8pt]
&-&\!
\big(\vert\tilde\mu\vert^2+ m_1^2\big) \,\,
\vert h_1\vert^2\,
-\big(\vert\tilde\mu\vert^2+ m_2^2\big)\,\,
\vert h_2 \vert^2\,
-\big(h_1\,h_2\,
B\mu
+h.c.\big)
-h_1^*\,\cD^2\,h_1 - h_2^*\,\cD^2\,h_2\nonumber
\eea
where
\bea\label{m1m2}
m_1^2
&=& M_s^2\,\Big( \vert\,a_1\,\vert^2-
a_2\Big) +\cO(M_s/M)
\nonumber\\
m_2^2
&=&
M_s^2\,\Big( \vert\,b_1\,\vert^2-
b_2\Big)+\cO(M_s/M)
\nonumber\\
B\mu
&=&
\tilde\mu\,M_s\,\Big(c_1-a_1-b_1\Big)
+\cO(M_s/M)\label{Bmu}
\eea

\medskip\noindent
The $\cO(M_s/M)$ corrections in (\ref{m1m2})
 are not shown explicitly  since they only renormalise
 $m_{1,2}$ and $B\mu$ which are anyway unknown parameters
of the MSSM.
In (\ref{finalscalar}) we  denoted
\bea\label{delta12only}
\delta_1& = & -\beta_1\,\alpha_0\,\gamma_0
+\gamma_1\,\beta_0\,\alpha_0\,
-\alpha_0\beta_0\gamma_0\,(a_1-b_1),
\quad\,\,
\delta_2= c_2+ 2 (a_1+b_1),
\eea
From  (\ref{finalscalar})  we
notice  the presence in the scalar potential
of three  contributions, all introduced by
our dimension-five operators. The
contributions proportional to $c_3$ in
 (\ref{finalscalar}) are due to $(H_1 H_2)^2$ in (\ref{lls})
and where discussed in \cite{Dine} (also \cite{Blum:2008ym,Strumia}; 
 for a review  see \cite{Brignole}).
The contribution proportional to $\delta_1$ in (\ref{finalscalar})
\bea
\big(\vert h_1\vert^2-\vert h_2\vert^2\big)\,
\big(h_1\,h_2+h.c.\big),
\eea
was introduced by the dimension-five operator in the last line of
(\ref{lls}). This is a new contribution to the scalar potential,
and  is vanishing if $\alpha_0=\beta_0=\gamma_0$.
An interesting feature of this new contribution to  the MSSM scalar
potential is that  its one-loop contribution to $h_{1,2}$
self-energy  remains soft (no quadratic divergences) despite
its higher dimensional origin\footnote{
One can ask  what happens to the value of
$\delta_1$ after one uses the remaining freedom of  rescaling the
chiral superfields in (\ref{lls}) as follows:
$ H_1\ra \,(1-a_1\,S)\,H_1;\,\,
H_2\ra \,(1-b_1\,S)\,H_2$.
Under such rescaling  $\beta_1\ra \beta_1-\beta_0\,b_1$,\,
$\gamma_1\ra\gamma_1-\gamma_0\,a_1$, see (\ref{definitions3}).
 Using the value of
$\delta_1$ in  (\ref{delta12only}) (now with $a_1=b_1=0$) and with these
new values of $\beta_1,\gamma_1$ one immediately sees that
$\delta_1$ is invariant/remains unchanged under this rescaling.}.

\subsection{Higgs mass corrections beyond the MSSM.}
\label{ss3}

Let us consider the implications of
(\ref{finalscalar}) for the Higgs masses.
The scalar potential is
\medskip
\bea\label{scalarV}
V&=&\tilde m_1^2\,\vert h_1\,\vert^2
+\tilde m_2^2\,\vert h_2\,\vert^2
+\Big( \,B\, \mu \,h_1 \,h_2+h.c.\Big)
+\frac{g^2}{8}\,\Big(\vert \,h_1\,\vert^2-\vert\,
h_2\,\vert^2\Big)^2
\nonumber\\[8pt]
&+& \Big(\vert\,h_1\,\vert^2
-\vert\,h_2\,\vert^2\Big)\,\Big(\eta_1 \,h_1\,h_2+h.c.\Big)
+
\Big(\vert\,h_1\,\vert^2+\vert\,h_2\,\vert^2\Big)\,
\Big(\eta_2 \,h_1\,h_2+h.c.\Big)
\nonumber\\[8pt]
&+&
\frac{1}{2}\,\Big(\,\eta_3\,(h_1\,h_2)^2+h.c.\Big)
\eea
where the  definition  of
 $\eta_{1,2,3}\sim 1/M$ can be read
from eq.(\ref{finalscalar}). We take for simplicity 
 $\eta_{i}$ real,  and  therefore
$\eta_3\geq 0$, $\vert\eta_2\vert\leq \eta_3/4$. 
Also
\bea
\tilde m_1^2&\equiv &m_1^2+\vert\tilde \mu\vert^2,\qquad\qquad
\tilde m_2^2\equiv m_2^2+\vert\,\tilde \mu\,\vert^2,\qquad\qquad
g^2\equiv g_1^2+g^2_2
\eea

\noindent
Consider quantum fluctuations
\bea
h_i=\frac{1}{\sqrt  2}\,(  v_i+\tilde h_i+i\tilde \sigma_i
),\qquad i=1,2
\eea

\medskip
\noindent
where $  v_{1,2}$ are the  minimum vev's of $V$.
Following the details presented in Appendix~D
and using the minimum conditions for $V$
one shows  that the Goldstone boson has $m_G=0$ and the
pseudoscalar  Higgs $(A)$ has a mass
\medskip
\bea\label{maprime}
m_A^2 &=&
-
\frac{1+u^2}{u}\,B\,\mu
+
\, \frac{u^2-1}{2 \,u}\, \eta_1 \,{  v}^2\,
-
\,\frac{1+u^2}{2\,u}\,\eta_2\,{  v}^2\,
-
\eta_3\,{  v}^2
\eea

\medskip\noindent
with the notation $u\equiv \tan\beta$ and $B\mu<0$.
Also
$  v_1=  v\cos\beta$, $  v_2=  v
\sin\beta$ and  $m_Z^2=g^2 \,{  v}^2/4$.
The masses of the CP even Higgs scalars $h,H$ are
(see also eq.(\ref{finalmh})):
\medskip
\bea\label{mh2}
 m_{h,H}^2&=&
\frac{1}{2}\Big[m_A^2+m_Z^2\mp\sqrt{w''}\Big]
\pm
\,\eta_1\,{  v}^2\,\sin 4\beta\,\,\frac{m_A^2}{\sqrt{w''}}
\nonumber\\[10pt]
&+&
{\eta_2\,{  v}^2\,\sin 2\beta}
\,\bigg[1\pm\frac{m_A^2+m_Z^2}{\sqrt{w''}}\bigg]+
\frac{\eta_3\,{  v}^2}{2}\,\bigg[1\mp
\frac{(m_A^2-m_Z^2)\,\cos^2 2\beta}{\sqrt{w''}}\bigg]
\eea

\medskip
\noindent
where the upper (lower) signs correspond to $h$ ($H$) respectively
and
\bea
w''\equiv
(m_A^2+m_Z^2)^2-4\,m_A^2\,m_Z^2\,\cos^2 2\beta
\eea

\medskip\noindent
For $\eta_2=\eta_3=0$ one finds from (\ref{mh2})
\bea
m_h^2 +  m_H^2 = m_A^2+m_Z^2
\eea
which is independent of $\eta_1$. Then $\eta_1$ does not
affect the relation among physical masses, which  is
consistent with  the result of Section~\ref{ss1},
where the last term in (\ref{lls}) responsible
 for $\eta_1$ term in $V$ could be removed
by a suitable field redefinition.

For $\eta_1=0$ the result in (\ref{mh2}) reproduces
 that in the first line of eq.(31) in
\cite{Dine}
\footnote{In the notation of
\cite{Dine}, our  $\eta_2=2\epsilon_{1 r}$ and $\eta_3=2\,\epsilon_{2 r}$
and $v$ has a different normalisation there.}.
In  the limit of large $\tan\beta$ with $m_A$ as a
parameter fixed at a value $m_A>m_Z$ one finds:
\medskip
\bea\label{rr1}
m_h^2&=&m_Z^2+\frac{4 m_A^2 \,{  v}^2}{m_A^2-m_Z^2}\,(\eta_2-\eta_1)
\,\cot  \beta\nonumber\\[10pt]
&-&\frac{4\,m_A^2\,m_Z^2}{m_A^2-m_Z^2}
\,\bigg[
1-\eta_3 \,{  v}^2\,\frac{m_A^4+m_Z^4}{2\,m_A^2\,m_Z^2 \,(m_A^2-m_Z^2)}
\bigg]\,\cot^2\beta+\cO(\cot^3\beta)
\eea
and
\bea\label{rr2}
m_H^2&=& m_A^2+\eta_3\,{  v}^2 +\frac{4
  \,(m_A^2\,\eta_1-m_Z^2\,\eta_2)\,{  v}^2}{m_A^2-m_Z^2}\,\cot\beta
\nonumber\\[10pt]
&+&
\frac{4\,m_A^2\,m_Z^2}{m_A^2-m_Z^2}\,\bigg[1-\eta_3\,{  v}^2
  \,\frac{m_A^4+m_Z^4}{2 \,m_A^2\,m_Z^2\,(m_A^2-m_Z^2)}
\bigg]\,\cot^2\beta+\cO(\cot^3\beta)
\eea
Therefore
\bea\label{rr3}
\delta
m_h^2&=&\frac{4\,m_A^2\,{  v}^2}{m_A^2-m_Z^2}\,(\eta_2-\eta_1)\,\cot\beta
+\cO(\cot^2\beta)\nonumber\\[10pt]
\delta m_H^2&=&
\eta_3\,{  v}^2 +\frac{4
  \,(m_A^2\,\eta_1-m_Z^2\,\eta_2)\,{  v}^2}{m_A^2-m_Z^2}\,\cot\beta
+\cO(\cot^2\beta)
\eea

\medskip\noindent
in agreement with \cite{Dine} for  $\eta_1=0$.
The above expansions for large $\tan\beta$ should be regarded with due
care, since in fact they are the results of a double series expansion,
in $\eta_i$ and $1/\tan\beta$.
Assuming $\eta_3=0$ (then $\eta_2=0$, too), 
the  term proportional to $\cot\beta$ in (\ref{rr1})
is larger than the sub-leading one ($\cot^2\beta$),
giving $ m_h^2-m_Z^2>0$ if $\vert\eta_1/g^2\vert\geq
1/(4\tan\beta)$. 
This bound is however outside the validity of the perturbative
 expansion  in $\eta_{1}$ as we shall see shortly\footnote{See
 the bounds from  (\ref{ssx3}) and discussion below.}, and then this
large $\tan\beta$ expansion is not useful.
If $\eta_{1,2}=0$ and  $\eta_3$ non-zero and positive 
then one could obtain $m_h > m_Z$ if the square bracket in 
(\ref{rr1}) is negative, which is more easily satisfied (for a small
$\eta_3$)
if $m_A$ is very close to $m_Z$, but then 
 the above large  $\tan\beta$ expansion is not reliable.

Let us therefore analyse  the validity of the corrections to
$m^2_{h,H}$ from eqs. (\ref{mh2}), (\ref{finalmh}), 
in the approximation used. 
For our perturbative  expansion in $\eta_i$ to be accurate
we  require that the $\eta_{i}$-dependent
entries in the mass matrix
 $\cM_{ij}$  (\ref{mij}) be much smaller than the
corresponding values of  these matrix elements
 in the  MSSM case. From this condition one finds\footnote{
One may find this condition too restrictive;
in principle it may not be necessary to impose the
leading $\eta_i\sim \cO(1/M)$ contribution to the mass matrix entries 
 be suppressed relative to the MSSM  
zeroth order and that one should instead ask that the $\cO(1/M)$
correction  dominate over the
higher order terms $\cO(1/M^2)$. However, at the quantitative level
this leads, for the present case,
  to results  which are similar or even stronger
(for example for $\eta_3$) than those derived here from comparing
the MSSM zeroth order against the $\cO(1/M)$ terms. (We thank
K.~Blum, Y.~Nir and G.G.~Ross for bringing this issue to our attention).}
\medskip
\bea\label{cond3}
&&\Big\vert
\, 3\, (\eta_1+\eta_2)\,  v_1^2+3 \,(\eta_2-\eta_1)\,  v_2^2
+2\eta_3\,  v_1\,  v_2\,
\Big\vert\ll
\frac{1}{2}\,g^2\,
  v_1\,v_2 
\nonumber\\[3pt]
&&
\Big\vert
\, 6 \,(\eta_2-\eta_1)\,  v_1\,  v_2+\eta_3\,  v_1^2
\,\Big\vert \ll 
\frac{1}{4}\,g^2\,\Big\vert\,
v_1^2-3 \,  v_2^2 \Big\vert
\nonumber\\[3pt]
&&
\Big\vert
\,6 \,(\eta_2+\eta_1)\,  v_1\,  v_2+\eta_3\,  v_2^2
\,\Big\vert \ll 
\frac{1}{4}\,g^2\,\,\Big\vert\,
3\,  v_1^2-  v_2^2 \Big\vert
\eea

\medskip\noindent
Similar conditions are derived
from the pseudoscalar Higgs/Goldstone bosons mass matrix 
elements $N_{ij}$ (\ref{nij}).
From these one can obtain some upper bounds  for each\footnote{Note that
 a non-zero $\eta_2$ requires nonzero $\eta_3$ since $\vert \eta_2\vert
  \leq \eta_3/4$.}
$\eta_i$; lower bounds on $\eta_{i}$ can be derived
 from the condition that
the contribution of each $\eta_i$ or combinations thereof
increase $m_h$ above $m_Z$ (to avoid the MSSM tree 
level bound $m_h\leq m_Z$).
If all these bounds on $\eta_i$ can be respected simultaneously,
then it is possible to obtain $m_h>m_Z$ in the approximation
considered.

Assuming $\eta_2=0$, then $m_h>m_Z$ is possible if one or both
eqs in  (\ref{ssx3}) are  respected.
One can show that for $1\leq   \tan\beta\leq 50$ and
$ m_A/m_Z \geq 1$ eq.(\ref{ssx3})
 has no solution for $\eta_{1}$; therefore
$\eta_1$ alone  cannot change the MSSM bound $m_h\leq m_Z$
within our approximation.
If  $1\leq m_A^2/m_Z^2 \leq 2.43$ there is
a  somewhat ``marginal''  solution for $\eta_3$ of (\ref{ssx3}),
with values of $m_A/m_Z$ close to unity and
with large $\tan\beta$ preferred, to enforce the 
``$\ll$'' inequalities in (\ref{cond3}), (\ref{ssx3}). For example,
for $m_A=m_Z$  and $\tan\beta=50$ the lower  bound on $\eta_3/g^2$ is
$\eta_3/g^2\geq 0.02$ while  $\eta_3/g^2\ll 0.25$ is also required; 
in this case, for
$\tan\beta=50$ the increase of $m_h$ relative to $m_Z$,
 $\delta_r=(m_h^2-m_Z^2)/m_Z^2$
equals $\delta_r=-100/2501+2\,\eta_3/g^2$. Therefore
 $\delta_r=12\%$ or $m_h\approx 102$ GeV
 if $\eta_3/g^2=0.08$, corresponding to $\eta_3=4.4
\times 10^{-2}$. Larger values for $m_h$ 
should  be regarded  with care, since would correspond to
cases when  $\ll$ of (\ref{ssx3}) is not
comfortably respected; if $\eta_3/g^2\approx 0.04$ then
 $\delta_r\approx 4\%$ or $m_h\approx 95$ GeV.
Further, if
 we now increase $m_A$ even by a small amount relative to $m_Z$,
 $m_A^2=1.5\, m_Z^2$\,  and $\tan\beta=50$ the lower  bound on
$\eta_3/g^2$ is  $0.118$,  difficult to 
comply by a good margin 
with an upper bound unchanged at $\eta_3/g^2\ll 0.25$.
Even so, then   $\delta_r=2\times 10^{-3}\%$ only,
if $\eta_3/g^2=0.118$ ($\eta_3=6.48\times 10^{-2}$),
therefore the increase of $m_h$ is negligible.
So far we took $\eta_2=0$; 
if we allow a non-zero value for $\eta_2$,
which also requires non-zero $\eta_3$,
their combined effect on increasing $m_h$
is not larger, and the above results remain valid.
Note also that for large $\tan\beta$ regions
 $1/M^2$-suppressed operators can be important 
and can affect the results \cite{Dine}.

From this analysis we see that  $\eta_{1}$ alone cannot
change   the MSSM tree level bound $m_h\!\leq\! m_Z$
within the approximation we discuss.
This  is consistent with  Section~\ref{ss1}, where it was shown that
the operator  which induced the $\eta_1$ term could be removed
by a general field redefinition of suitable 
coefficients\footnote{To see this
 one can also start from (\ref{lls}) and perform
a ``smaller'' version of redefinition (\ref{tra}), with $\rho_F\!=\!0$.}.
However,
$\eta_3$ can increase $m_h$ to values 
$\approx 95-100$ GeV  if $m_A\approx m_Z$, with the higher values close
to the limit of our approximation.
Therefore it is  the susy breaking term
associated to $(H_1\,H_2)^2$ that could
relax the MSSM tree level bound.
This increase brings a  small improvement. 
To conclude, adding the quantum corrections
is still needed \cite{Dine} to bring $m_h$ above the 
LEP II bound of 114 GeV \cite{higgsboundLEP}.

These findings show that the MSSM Higgs sector is rather
stable under the addition of D=5 operators, in the approximation
 we considered (expansion in $1/M$) of integrating out a
massive singlet or 
a pair of massive $SU(2)$ doublets which generated the $\eta_{1,2,3}$ 
contributions.
If $M$ is low-enough, the approximation used of
integrating out these massive fields becomes
unreliable, and one should re-compute the full spectrum with all
fields  un-integrated out. Then  the quartic interactions
that the initial massive fields  brought can be larger or of similar
order (rather than corrections)  to their  MSSM counterparts,
and can change the above conclusions.

\section{Conclusions}

In this work we considered a natural
 extension of the MSSM by the addition
of R-parity conserving dimension-five operators and analysed some of
their implications. As we showed, such operators are a common
presence in effective theories, generated  by integrating out
massive singlets and  $SU(2)$ doublets superfields.
 As it turns out,
not all these higher dimensional operators  are independent. We
presented a  method which employs general, spurion dependent field
transformations to identify  the minimal, irreducible set of such
operators  that one has  beyond the MSSM. This is done by using
field redefinitions suitably chosen to remove some of the
``redundant'' operators, up to renormalisations of the $\mu$-term
and of the soft terms. As a result, the low energy effective theory
has the advantage of a smaller number of  couplings (i.e.
parameters) and its study is simplified. The method can be applied
to other, more general models too.

The minimal set of D=5 operators can  be reduced further provided
that appropriate relations exist between the original couplings of
the dimension-five operators and the usual MSSM Yukawa couplings.
Such relations are expected to exist in the original Lagrangian to
avoid FCNC constraints. In this case, at order $1/M$, one is left
 with $(H_1\,H_2)^2$ and three additional Higgs-dependent
D-terms (\ref{lastL}), together with associated,
spurion-induced supersymmetry  breaking terms of a particular type.
 The superpotential
couplings and their associated soft terms acquire, already at the
classical level, nontrivial renormalisations,  which depend on the
scale $M$ of the higher dimensional operators. If our FCNC ansatz is
imposed only for the first two generations, quartic terms in the
superpotential $Q_3U^c_3 Q_3D^c_3$ and $Q_3U^c_3 L_3E^c_3$ are also
irreducible.

The  dimension-five Higgs-dependent D-terms leftover affect the
couplings of the model MSSM$_5$. In components, these terms contain
``wrong''-higgs (susy breaking) Yukawa  couplings. These are also
known to be generated in the MSSM alone at one-loop level by
integrating out massive squarks; our new contributions can be
significant if the new physics is not far above LHC energies.
 The combined effect of the two sources for
these couplings brings  a
 $\tan\beta$-enhancement of the mass of the bottom  quark.
Even more interesting are supersymmetric couplings of type
quark-quark-squark-squark and also quark-quark-slepton-slepton, that
are  also generated from the aforementioned D-term operators of
dimension five and/or by the quartic superpotential couplings if the
FCNC ansatz is made only for the first two generations. These
couplings, although suppressed  by $1/M$ can contribute
significantly, for the case of the third generation, to the process of 
squark production. This contribution competes with that of the similar
process coming from the MSSM at the tree level. This is
phenomenologically important since direct squark production can be a
first indication of supersymmetry at the LHC and this process is
significantly enhanced in the model  we discussed.

We also addressed the effects that dimension-five operators
have on the Higgs sector.
We included all possible contributions of the operators that
can be in general present, due to
$\cO_1=A(S,S^\dagger) D^\alpha\,[ B(S,S^\dagger) H_2\,e^{-V_1}\,]
D_\alpha\,[\,\Gamma(S,S^\dagger)
\,e^{V_1} H_1\,]$ and  $\cO_2=\lambda_H(S)(H_1\,H_2)^2$.
The analysis showed that
the MSSM  tree level bound $m_h\leq m_Z$
cannot easily be lifted by $\cO_{1,2}$ and their associated susy breaking
terms.  In the case of $\cO_1$ this is due to the fact that this
is ultimately a ``redundant'' operator and can be removed by a
field redefinition, as showed in Section~\ref{ss1}.
$\cO_1$ brings  ultimately only
a renormalisation of the soft terms and of supersymmetric $\mu$-term.
Within the approximation used,   the non-susy part of
$\cO_2$ can bring (somewhat close to
 the limit of validity of our approximation), 
 an increase of $m_h$ to $m_h\approx 95-100$ GeV,
while in that case $m_A\approx m_Z$.
 This shows that the MSSM Higgs sector is rather
stable under the addition of D=5 operators, in the approximation
 we considered.
This result for the Higgs sector is somewhat expected
in an effective theory where additional higher dimensional operators
can only bring small  corrections
to current relations among physical observables of the initial model.
Therefore quantum corrections are still needed to increase
$m_h$ above the LEPII bound of $114$ GeV.

In  conclusion,  the natural extension of the MSSM with the minimal, 
irreducible set of R-parity conserving dimension-five 
 operators  that we identified,  provides a
 consistent and very interesting framework
 for future detailed phenomenological  studies.  
The method presented to identify the minimal set of  
these operators beyond the MSSM  is general and
can be applied to sets of operators of higher dimensions
and/or of different  symmetries.

\bigskip
\section*{Acknowledgements}

This work was partially supported by ANR (CNRS-USAR) contract
05-BLAN-007901,   INTAS grant 03-51-6346,
EC contracts MRTN-CT-2004-005104, MRTN-CT-2004-503369,
MRTN-CT-2006-035863,  CNRS PICS
\#~2530,  3059, 3747, 4172,  and  European Union Excellence Grant
MEXT-CT-2003-509661. E.D. and D.G. thank the CERN Theory Group
and D.G. also thanks the Theory Group at \'Ecole
Polytechnique Paris, for their kind hospitality and support during 
their visits, at an early stage of  this work.
The authors thank K.~Blum, Y.~Nir and G.~G.~Ross for interesting
discussions on related topics.

\section{Appendix}

\def\theequation{A-\arabic{equation}}
\def\thesubsection{A}
\setcounter{equation}{0}

\subsection{Integrating out massive
superfields:  no gauge interactions present.}
\label{appendixA}

In this appendix we examine different methods of integrating
out high scale physics and confirm their equivalence,
 by showing that the same low energy effective Lagrangian is obtained.
We ignore gauge interactions, included in Appendix~\ref{appendixB}. 
We  find that integrating out massive states generates
in the effective action
and in the lowest order in the high scale,
a (classical) wavefunction renormalisation while in the next order
higher dimensional operators emerge.
Operators like $\Phi_2 \,D^2\Phi_1$ emerge,  which  in the
presence of gauge interactions
becomes  $\Phi_2\,e^{-V}\,D^2\,e^V\Phi_1$, studied in the text, 
Section~\ref{ss1}.
Let us start with a 4D renormalisable model (with $M\gg m$)
\bea\label{l0}
\cL_1=\int d^4\theta\, \Big[\Phi^\dagger \Phi +\chi^\dagger \chi\Big]
+\bigg\{
\int d^2\theta \bigg[\frac{M}{2}\,\chi^2+m\,\Phi\,\chi+
\frac{\lambda}{3}\,\Phi^3\bigg]+h.c.\bigg\}
\eea
\medskip
\noindent
With a transformation
 $\Phi\equiv(\cos\theta \,\Phi_1 -\sin\theta\,\Phi_2)$ and
$\chi\equiv (\sin\theta \,\Phi_1 +\cos\theta\,\Phi_2)$ one finds
\bea
\cL_1=\!\int d^4\theta\, \Big[\Phi_1^\dagger \Phi_1 +\Phi_2^\dagger
\Phi_2\Big]\! +\bigg\{
\int d^2\theta \bigg[\frac{m_1}{2}\,\Phi_1^2+\frac{m_2}{2}\,\Phi_2^2
+\frac{\lambda}{3}\,(\cos\theta \Phi_1 -\sin\theta\,\Phi_2)^3
\bigg]\!+\! h.c.\!\bigg\}
\eea
where
\bea
m_1 &=& \frac{M}{2}\,\Big(1-(1+4 m^2/M^2)^{1/2}\Big)
=-\frac{m^2}{M}\,\Big(1-\frac{m^2}{M^2}\Big)+\cdots
\nonumber\\[4pt]
m_2&=&\frac{M}{2}\,\Big(1+(1+4 m^2/M^2)^{1/2}\Big)
=M\,\Big(
1+ \frac{m^2}{M^2}+\cdots\Big),
\eea

\medskip\noindent
so $\Phi_2$ is the massive field.
We can now integrate out $\Phi_2$ via its equations of motion
\medskip
\bea
-\frac{1}{4} \overline D^2  \Phi_2^\dagger+m_2
\,\Phi_2-\lambda\sin\theta\,
\,\big(\Phi_1\,\cos\theta-\Phi_2\sin\theta\big)^2=0
\eea
with the solution
\bea
\Phi_2=\frac{\lambda}{m_2} \,\cos^2\theta\,\sin\theta\,\,
 \Phi_1^2
-\frac{\lambda^2}{4 m_2^2}\,\sin^3 2\theta\,\Phi_1^3
+
\frac{\lambda}{4\,m_2^2}\cos^2\theta\,\sin\theta\,
\overline D^2\Phi_1^{\dagger\, 2} +\cO(1/M^3).
\eea

\medskip\noindent
Keeping the lowest, dimension-five operators of $\cL_1$, we have
\medskip
\bea\label{m1}
\cL_1=\int d^4\theta\, \Phi_1^\dagger \Phi_1  +\bigg\{
\int d^2\theta \bigg[\frac{-m^2}{2 \,M}\,Z\,\Phi_1^2
+\frac{\lambda}{3}\,Z^{3/2}\,\Phi_1^3 -\frac{m^2\lambda^2}{2 M^3}
\Phi^4_1 \bigg]+h.c.\bigg\}+\cO(1/M^4),
\eea
where
\bea
Z=1-\frac{m^2}{M^2}+\cO(1/M^4)
\eea
As expected, we find that at low energies $(\ll M)$
a higher dimensional operator $\Phi_1^4$ emerges, suppressed by the scale
$M$ of ``new physics'' represented by the massive state $\chi$.
Other higher dimensional operators are present beyond that of
$\cO(1/M^3)$ shown, and these include higher  derivative
operators involving
$\overline D^2\Phi_1^{\dagger \,2}$.
As expected, in the low energy limit,  the initial 4D renormalisable
theory appears as an effective field theory valid below the scale $M$.

There is another, equivalent way to
analyse the  Lagrangian in (\ref{l0}) in the low energy limit,
which  illustrates further the emergence
of higher dimensional operators.
Start again with eq.(\ref{l0}), which gives the following eq of motion
for the massive field $\chi$:
\bea\label{sss1}
0&=& {\overline D}^2\chi^\dagger -4\,(M\,\chi+m\,\Phi)
\eea
with an iterative solution
\bea\label{eq_sol}
\chi
&=&
\frac{1}{M}\Big[-m\,\Phi-\frac{m}{4M}\,{\overline D}^2\Phi^\dagger
+\frac{1}{16}\,\frac{-m}{M^2}\,{\overline D}^2 \,D^2\Phi
-\frac{m}{64\, M^3}\,\overline D^2\, D^2\,\overline D^2\Phi^\dagger
+\cdots\Big]\quad
\eea

\bigskip
\noindent
Using this solution
in original $\cL_1$ of (\ref{l0}), one finds
\bigskip
\bea\label{ff1}
\cL_1&=&
\int d^4\theta\,\bigg\{
\bigg[\,1+\frac{m^2}{M^2}\bigg]
\,\Phi^\dagger\Phi+
\frac{m^2}{8\,M^3} \,\Big[\Phi\,D^2\,\Phi+h.c.\Big]
+\frac{m^2}{16\, M^4}\,(\overline D^2\Phi^\dagger)\, (D^2 \Phi)
\bigg\}
\nonumber\\[10pt]
&+&
\bigg\{
\int d^2\theta
\,\,\bigg[\frac{-m^2}{2 M}\,\Phi^2+ \frac{\lambda}{3}\,\Phi^3\bigg]
+h.c.\bigg\}+\cO({1}/{M^5})
\eea

\medskip\noindent
After an appropriate re-scaling
\bigskip
\bea\label{ff2}
\cL_1&=&
\int d^4\theta\,\bigg\{
\,\Phi^\dagger\Phi+
\frac{m^2}{8\,M^3}\, \,\Big[\Phi\,D^2\,\Phi+h.c.\Big]
+\frac{m^2}{16\, M^4}\,(\overline D^2\Phi^\dagger)\, (D^2 \Phi)
\bigg\}
\nonumber\\[10pt]
&+&
\bigg\{
\int d^2\theta
\,\,\bigg[\frac{-m^2}{2 M}\,Z\,\Phi^2+
\frac{\lambda}{3}\,Z^{3/2}\,\Phi^3
\bigg]
+h.c.\bigg\}+\cO({1}/{M^5})
\eea

\medskip\noindent
where $Z=1/(1+m^2/M^2)$.
After\footnote{
Using $\overline D^2 D^2 =-16 \Box$ we find a $-\Phi^\dagger
\Box\Phi$ term; the metric is $(+,-,-,-)$.}
integrating out a massive superfield $\chi$,
higher dimensional derivative operators were generated.
These are suppressed by $M$,
below which only an  effective theory (\ref{ff2}) applies.
Since the presence of massive states in high scale theories is
usually expected, the conclusion is that this type of operators are
a generic presence at low energies.
There are no ghosts in  $\cL_1$ of (\ref{ff2}) as long as
one keeps all terms in the series\footnote{This is true because the
 original theory (\ref{l0}) had no ghosts;
for a detailed
discussion see \cite{Antoniadis:2007xc,Antoniadis:2006pc}.} (\ref{eq_sol}).
Once we truncate this series to a given order, such states can be
generated, as a signature of the fact that the UV of the theory is
unknown. Finally, in order $1/M^2$ the only effect of the
massive state is
a wavefunction renormalisation which depends on high scale $M$.

From this stage there are two approaches one can adopt to
continue from eq.(\ref{ff2}).

\noindent
I). In the first approach  one  sets {\it ``onshell''} the
higher dimensional operator, using the equations of 
motion\footnote{For an application see \cite{Giudice:2007fh}.}, see
\cite{Georgi:1991ch,Politzer:1980me,Arzt:1993gz}; if one
adheres to this procedure, the eq of motion
\bea\label{ggg1}
\overline D^2 \Phi^\dagger =
-\frac{4 m^2}{M}\,\Phi+4\,\lambda\,\Phi^2+\cO(1/M^2)
\eea
can be used back in (\ref{ff2}); the new Lagrangian so
 obtained will contain
a term $\Phi\Phi^{\dagger 2}$ which can be removed by a suitable
shift
\bea
\Phi=\tilde\Phi-\frac{\lambda \,m^2}{2\,M^3}\,{\tilde\Phi}^{2}
\eea
to finally find
\bea\label{ggg2}
\cL_1&=&
\!\!\!
\int d^4\theta \,\,
\tilde\Phi^\dagger\tilde\Phi
\\[10pt]
&+&\!\!\!\bigg\{
\int d^2\theta\, \bigg[-\frac{m^2}{2M}\,Z\,\tilde\Phi^2
+
\frac{\lambda}{3}\,\tilde\Phi^3\,\Big(\,1-\frac{3}{2}
\frac{m^2}{M^2}
\,\Big)-
\frac{\lambda^2\,m^2}{2\,M^3}\,\tilde\Phi^4\bigg]\!+\!h.c.\!\bigg\}
+\cO\Big(\frac{1}{M^4}\Big)\nonumber
\eea

\medskip\noindent
where $Z=1/(1+m^2/M^2)$.
In the approximation  $\cO(1/M^4)$
this Lagrangian coincides with that of (\ref{m1}), where
a different method was used. This confirms that
 setting the higher derivative
operators ``onshell'' via equations of motion is a correct procedure,
 within the approximation considered.
We again obtained a higher dimensional operator
and a scale dependence acquired {\it classically}
by the couplings of the low energy effective theory\footnote{
To the  next order, in  (\ref{ggg2}) one has extra D terms
$(m^2\,\lambda^2/M^4)\,\tilde\Phi^2\,\tilde\Phi^{\dagger 2}$
and F terms $(39\, m^4/(8 M^4))\, \Phi^3$.}.

\noindent
II). Finally let us now take the second approach to continue
from the Lagrangian in
(\ref{ff2}). This will provide another check that
setting  onshell the higher derivative operators as done
above in I) is indeed 
a correct procedure. In  eq.(\ref{ff2}) proceed to redefine the fields,
to eliminate the $\Phi D^2\Phi$ term. We use a field redefinition
\bea\label{hh1}
\Phi=   \Phi'+c\,\overline D^2\Phi^{' \dagger}
\eea
where the dimensionful coefficient $c$  is found from
the requirement that the coefficient of
$\Phi D^2\Phi$ vanish in the new Lagrangian.
This gives $c=-m^2/(8 M^3)$ and the Lagrangian in (\ref{ff2}) becomes
after some calculations

\bea\label{hh2}
\cL_1&=&\int d^4\theta \,\,\Big[
\Phi^{' \dagger} \,\Phi'+\frac{m^2\,\lambda}{2
  \,M^3}\,\big(\Phi^{' 2}\,\Phi^{' \dagger}+h.c.\big)\Big]\nonumber\\[10pt]
&+&
\bigg\{\!
\int d^2\theta\,\,
\Big[-\frac{m^2}{2 \,M}
  \,Z\,\Phi^{' 2}+
\frac{\lambda}{3}\,Z^{3/2}\,\Phi^{' 3}\Big]+h.c.\bigg\}
+\cO(1/M^4)
\eea

\bigskip\noindent
After a shift
$\Phi'= \tilde \Phi-{m^2\,\lambda}/(2\,M^3)\,\tilde\Phi^{2}$
we obtain a low energy Lagrangian identical to that in (\ref{m1}),
 (\ref{ggg2}).  This result shows
that  the three approaches to integrating out the
effects of high scale physics ($\chi$), using $a)$
eqs.(\ref{l0}) to (\ref{m1}), or $b)$
 setting the higher dimensional derivative operators ``onshell''
eqs.(\ref{sss1}) to  (\ref{ggg2}), and finally $c)$  using field
re-definitions (\ref{hh1}), are equivalent  to the lowest
order studied. The approaches gave in all cases
 the same spectrum and couplings, and checked explicitly that
setting onshell the higher derivative operators is correct
in the approximation considered.
To the lowest order in $1/M$ only a
 wavefunction renormalisation was
introduced by integrating out massive states,
which classically renormalise low energy couplings.
Higher dimensional operators were generated in
the next order in $1/M$.

\def\theequation{B-\arabic{equation}}
\def\thesubsection{B}
\setcounter{equation}{0}

\subsection{Integrating out massive
superfields: gauge interactions present.}
\label{appendixB}

Here we show
how  all  dimension-five operators of $\cL^{(5)}$ of
eq.(\ref{ll5}) in Section~\ref{ss1}  are generated, 
 and discuss in particular  $\Phi_2\,e^{-V}\,D^2\,e^V\,\Phi_1$.
 This appendix also extends the analysis
 in Appendix~\ref{appendixA} where a similar
$\Phi\,D^2\Phi$ was shown to
 arise, in the absence of gauge interactions.
Consider the Lagrangian of a N=1 supersymmetric
non-Abelian gauge theory\footnote{For the link to the MSSM,
replace $V\ra V_1\equiv g_2 V_w^i\sigma^i-g_1 V_Y$
with $V_w, (V_Y)$ the $SU(2)$, ($U(1)_Y$) gauge fields respectively;
also $\Phi_2\ra H_2^T\,(i\sigma_2)$, $\Phi_1\ra H_1$ with $\Phi_3$
$(\Phi_4)$ with same quantum numbers to $\Phi_1$ ($\Phi_2$) and
$(i\sigma_2)\exp( -\Lambda)=\exp(\Lambda^T)\,(i\sigma_2)$,
then 
$\Phi_2\,e^{-V}\,\Phi_2^\dagger\ra  H_2^\dagger\,e^{V_2}\,H_2$,
 with $V_2\equiv g_2 V_w^i\sigma^i+g_1 V_Y$.}
\medskip
\bea\label{LO}
\cL_2&=&\int d^4\theta
\,\,\Big[
\,\Phi_1^{ \dagger}\,\,e^{V}\,\Phi_1
+\,\Phi_3^{ \dagger}\,\,e^{V}\,\Phi_3
+\,\Phi_2\,\,e^{-V}\,\Phi_2^\dagger
+\,\Phi_4\,\,e^{-V}\,\Phi_4^\dagger\,\Big]
\nonumber\\[6pt]
&+&
\int\,d^4\theta\,\,
\Big[\nu_1\,\Phi_1^{ \dagger}\,\, e^{V}\, \Phi_3+
\nu_2\,\Phi_4\, e^{- V}\, \Phi_2^\dagger+h.c.\Big]
\nonumber\\[6pt]
&+&
\int d^2\theta
\,\,\Big[\,\mu\,\Phi_1\,\Phi_2+M\,\Phi_3\,\Phi_4
+\cW'\Big]+h.c.
\eea

\medskip
\noindent
where $M\gg \mu$ and with the notation $V\equiv (V_\mu,
\lambda, D/2)$ in the Wess-Zumino gauge. For generality and for
phenomenological applications we can allow the presence of another
higher dimension term
 $\cW'= \int d^2\theta \,\xi'\,(\Phi_1\,\Phi_2)^2$,
where we assume  $\xi'\sim \cO(1/M)$; ($\cW'$ can
 be generated by integrating out a singlet).
The equations of motion for massive $\Phi_{3,4}$ give
\medskip
\bea
-\frac{\nu_1}{4}\,\oD^2\,\Big(\,\Phi_1^\dagger\,e^{V}\Big)-\frac{1}{4}
\,\oD^2\Big(\Phi_3^\dagger \,\,e^{V}\Big)+M\,\Phi_4&=&0
\nonumber\\[8pt]
-\frac{\nu_2}{4}\,\oD^2\,\Big(\,e^{-
V}\,\Phi_2^\dagger\,\Big)-\frac{1}{4} \oD^2\Big(\,e^{-
V}\,\Phi_4^\dagger\,\Big)+M\,\Phi_3&=&0 \label{integratephi}
\eea

\bigskip
\noindent
As in previous section
 we use these equations to integrate out the massive fields
$\Phi_{3,4}$ to find
\bigskip
\bea\label{lag4}
\cL_2&=&\int d^4\theta
\,\,\Big[
\,\Phi_1^{ \dagger}\,\,e^{V}\,\Phi_1
+\,\Phi_2\,\,e^{- V}\,\Phi_2^\dagger
+\Big(
\,\frac{\nu_1\,\nu_2}{4} \, \,\xi\,\,
\,\Phi_1^\dagger\,\,e^{V}\,\oD^2\,e^{-V}\,\Phi_2^\dagger
+h.c.\Big)
\,\Big]
\nonumber\\[8pt]
&+&
\int d^2\theta
\,\,\Big[\,\mu\,\Phi_1\,\Phi_2+ \xi'\,(\Phi_1\,\Phi_2)^2\Big]
+h.c.+\cO(1/M^2),
\qquad\qquad
\xi\equiv \frac{1}{M}
\eea

\bigskip
\noindent
where we ignored higher  orders in  $1/M$.
Again, higher dimensional operators were generated by integrating out
massive superfields $\Phi_{3,4}$, as expected in the low energy
effective action.
Before a  detailed analysis of (\ref{lag4}), let
us set onshell the first dimension-five operator in 
(\ref{lag4}) by using the equations of motion for $\Phi_{1,2}$:
\bea\label{eqm1}
D^2 \,\big[\,e^{V}\, \Phi_1\,\big]
=4 \,\mu\,\Phi_2^\dagger, \qquad\qquad
\overline D^2 \,\big[\, e^{- V}\, \Phi_2^\dagger\,\big] =4\,\mu\,\Phi_1
\eea
We insert these in (\ref{lag4}), then rescale
$\Phi_i\ra \Phi_i' \,(1-\mu\,\,\nu_1\,\nu_2\,\xi/2)$, $i=1,2$,
to find:
\medskip
\bea\label{adasd}
\cL_2&=&\int d^4\theta
\,\,\Big[
\,\Phi_1^{ \dagger}\,\,e^{V}\,\Phi_1
+\,\Phi_2\,\,e^{-V}\,\Phi_2^\dagger
\,\Big]
\nonumber\\[8pt]
&&\qquad +\,
\int d^2\theta \,\,\Big[\,\mu\,(1- \mu\,\nu_1\,\nu_2 \,\xi)\,\,
\Phi_1\,\Phi_2+ \xi'\,(\Phi_1\,\Phi_2)^2 \Big]+h.c.+\cO(1/M^2),
\eea

\medskip
\noindent In conclusion,  the supersymmetric higher dimensional
operator  (generated by integrating out massive superfields), when set
on-shell, produced  in the leading order (in 1/M) only wavefunction
renormalisation. The D=5 D-term operator in (\ref{lag4}) was
studied in Section~\ref{ss1}.

 If the superpotential in (\ref{LO}) also contains 
trilinear  couplings of the heavy doublets $\Phi_{3,4}$ to 
the quarks
\begin{equation}
\Delta {\cal L}_2 \ = \ \int d^2 \theta \Big[ 
Q\, \sigma_u  U^c\,\Phi_4 + 
Q\, \sigma_d  D^c \,\Phi_3 + 
L\, \sigma_e  E^c \,\Phi_3  \Big]+h.c. \ , \label{others}
\end{equation}
then they  change the rhs of (\ref{integratephi}) by extra terms and
then new higher dimensional  operators are also generated in the low
energy effective action in addition to the first one in (\ref{lag4}).
More precisely, the Lagrangian in (\ref{lag4}) 
 acquires a correction
\begin{eqnarray}\label{B7}
\Delta \cL_2'&= & \, - \frac{1}{M} \int d^4\theta \,\,
\Big[\nu_1\, \Phi_1^\dagger\,e^{V}\,Q\,  \sigma_u U^c\,  
+ \nu_2\,(Q\,  \sigma_d D^c) \, e^{-V} \,\Phi_2^\dagger \, 
+ \nu_2\,(L\, \sigma_e E^c) \, e^{-V} \,\Phi_2^\dagger \,+
h.c.\Big] \nonumber\\[5pt]
   &+& 
\frac{1}{M}\int d^2 \theta \Big[
(Q\sigma_u U^c) ( Q\sigma_d D^c ) 
+
(Q\sigma_u U^c) (L\,\sigma_e E^c)\Big]+h.c. \ ,
\end{eqnarray}
where $\sigma_{u,d,e}$ are 3x3 matrices in the
family space. This gives one possible origin of the  D=5 operators
analysed in Section~\ref{ss1}.
Eq.(\ref{B7}) generates tree-level ``wrong-Higgs'' couplings 
and  fermion-fermion-sfermion-sfermion
 couplings,   discussed in Section~\ref{ss1} and \ref{ss1p}. 
The structure of the couplings in (\ref{B7})
 also motivates the ansatz made 
in Section~\ref{ss1p}. Eq.(\ref{ansatz0}) would be obtained if
$\sigma_F\propto \lambda_F$, $F:U,D,E$, which could 
eventually be enforced by family symmetries.

In the remaining part of this section we  present the
general  {\it offshell} form of $\cL_2$  of  (\ref{lag4}).
Using now this form, we check again
 that the  higher dimensional (derivative)
operator in (\ref{lag4}) brings a  wavefunction
renormalisation only, in the absence of other interactions
coupled to $\Phi_{1,2}$ (like trilinear terms).
After a long calculation, one obtains the
 offshell form\footnote{We use
$-4\,\psi_2\,\cD_\mu\,\cD^\mu\,\psi_1
=
-4\,\psi_2\,
[\sigma^\nu\,\overline\sigma^\mu-2\,i\,\sigma^{\mu\nu}]
\cD_\nu\,\cD_\mu\,\psi_1
%\nonumber\\[10pt]
=
-4\,\psi_2\,\sigma^\nu\,\overline\sigma^\mu\,\cD_\nu\,\cD_\mu
\,\psi_1+4\,\,\psi_2\,\sigma^{\mu\nu}
\,F_{\mu\nu}\,\psi_1$
and the first term
in the rhs is that entering the final
expression of $\cL_2$. Here $F_{\mu\nu}=\partial_\mu V_\nu/2-
\partial_\nu V_\mu/2+i\,[V_\mu/2,V_\nu/2]$.}
\bigskip
\bea\label{offshell}
\cL_2&=&
-\, \phi_1^*\,\cD_\mu\cD^\mu \phi_1
+i\,
\overline\psi_1\, \overline\sigma^\mu\,\cD_\mu\,\psi_1
-\frac{1}{\sqrt 2}\,\Big[
\overline\psi_1\,\overline\lambda\,\phi_1
+h.c.
\Big]
+\phi_1^*\,\frac{D}{2}\,\phi_1
+\vert F_1\vert^2
\nonumber\\[7pt]
&-&
\phi_2\,\cD_\mu\cD^\mu \phi_2^*
+i\,
\psi_2\,\sigma^\mu\,\cD_\mu\,\overline\psi_2
+\frac{1}{\sqrt 2}\,\Big[
\phi_2\,\overline\lambda\,\overline\psi_2
+h.c.
\Big]
-\, \phi_2\,\frac{D}{2}\,\phi_2^*+\vert F_2\vert^2
\nonumber\\[7pt]
&+&
\frac{1}{4} \,\nu_1^*\, \nu_2^* \,\,\xi\,\,\bigg\{
4\, \Big[\,F_2 \,\cD_\mu\,\cD^\mu\,\phi_1+
\phi_2\,\cD_\mu\,\cD^\mu\,F_1
\Big]
+
2\sqrt 2\,i\,\,\Big[\psi_2\,\sigma^\mu\,
\overleftarrow\cD_\mu\,\overline\lambda\,\phi_1+
\phi_2\,\overline\lambda\,\,\overline\sigma^\mu\,\cD_\mu\,\psi_1\Big]
\nonumber\\[7pt]
&+&
2 \,(\phi_2\, D\, F_1-F_2\, D \,\phi_1)
-
2\sqrt 2 \,\,\Big[
\psi_2\,\lambda\,F_1-F_2\,(\lambda\,\psi_1)\Big]
-2\,\phi_2\,(\overline\lambda\,\overline\lambda)\,\,\phi_1
\nonumber\\[7pt]
&-&
4\,\psi_2\sigma^\nu\,\overline\sigma^\mu\,\cD_\nu\,\cD_\mu\psi_1
\bigg\}+
\mu\,\Big[\phi_1\,F_2+F_1\,\phi_2-\psi_1\,\psi_2\Big]+
{\cW'}{\big\vert}_{\theta^2}+h.c.+
\cO(1/M^2)
\eea

\medskip
\noindent
where
\bea
\cW'\big\vert_{\theta^2}= \xi'\,\Big[-
\big(\phi_1\psi_2+\psi_1\phi_2\big)^2+
2\,\big(\phi_1\phi_2\big)\,\big(\phi_1\,F_2
+F_1\phi_2-\psi_1\psi_2\big)\Big]
\eea
and with
\bea
\cD_\mu=\partial_\mu + i\,\,\frac{V_\mu}{2},\qquad
\overleftarrow\cD_\mu=\overleftarrow
\partial_\mu- i\,\,\frac{V_\mu}{2},\qquad
\eea

\medskip
\noindent
The first and second lines in (\ref{offshell})
are obtained from the first and second
terms in (\ref{lag4}) respectively; the  h.c. applies to all terms
in the last three lines of (\ref{offshell}).
In the offshell component form of the Lagrangian notice
we have an interesting tensor coupling
$\psi_2\,\sigma^\nu\,\overline\sigma^\mu\,\cD_\nu\cD_\mu\,\psi_1$
in spite of the minimal gauge coupling in (\ref{LO})
and this arises from a coupling $\psi_2\,\sigma^{\mu\nu}
\,F_{\mu\nu}\,\psi_1$ coming from the third term in the first line
of  (\ref{lag4}), see also the previous footnote
\footnote{
This  coupling is not present in the onshell form of the action,
see also \cite{Ferrara}.}.
This coupling could
 be relevant for tree level calculations of the Feynman
diagrams. Next we eliminate the auxiliary fields
$F_{1,2}$ using their equations of motion
\medskip
\bea
F_1^*
&=&
-\phi_2\,\Big(\mu+2\,\xi'\,(\phi_1\,\phi_2)\Big)
+
\frac{1}{4}\,\nu_1^*\,\nu_2^*\,\xi\,
 \,\Big(-4\,\phi_2 \overleftarrow\cD_\mu
\overleftarrow\cD^\mu
-4\,\,\phi_2\,\frac{D}{2}+2\sqrt 2 \,\psi_2\,\lambda\Big)
\nonumber\\[8pt]
F_2^*&=&
-\phi_1\,\Big(\mu+2\,\xi'\,(\phi_1\,\phi_2)\Big)
 +
\frac{1}{4}\,\nu_1^*\,\nu_2^*\,\xi\,
\,\Big(-4\,\cD_\mu\cD^\mu\,\phi_1
+4\,\,\frac{D}{2}\,\phi_1-2\sqrt 2 \,\lambda\,\psi_1\Big)
\eea

\medskip\noindent
In the terms proportional to $\xi$ in $\cL_2$ we can replace
the derivatives of the
fermions by their equations of motion, since the error would be
of higher order. We use there
\medskip
\bea
i\,\overline\sigma^\mu\,\cD_\mu\psi_1&=&
\mu\,\overline\psi_2+\frac{1}{\sqrt
  2}\,\,\overline\lambda\,\phi_1+\cO(\xi)
\nonumber\\[10pt]
-i\,\psi_2\,\sigma^\mu\,\overleftarrow
\cD_\mu&=&\mu\,\overline\psi_1-\frac{1}{\sqrt
  2}\,\,\phi_2\overline\lambda+\cO(\xi)
\eea
We then rescale the scalars and Weyl fermions
and after neglecting terms $\cO(\xi\,\xi')$
we obtain the onshell Lagrangian
\medskip
\bea\label{qqqq}
\cL_2&=&
-\,\phi_1^\dagger\,\cD^2\,\phi_1+
i\,\overline\psi_1\,\overline\sigma^\mu\,\cD_\mu\,\psi_1
- \frac{1}{\sqrt 2}\,\Big[
\overline\psi_1\,\overline\lambda\,\phi_1+h.c.\Big]
+\,\phi_1^\dagger\,\frac{D}{2}\,\phi_1
\nonumber\\[7pt]
&-&
\phi_2\,\cD^2\,\phi_2^\dagger +
i\,\psi_2\,\sigma^\mu\,\cD_\mu
\overline\psi_2+
\,\frac{1}{\sqrt
  2}\Big[\,\phi_2\,\overline\lambda\,\overline\psi_2+h.c.\Big]
-\,\phi_2\,\frac{D}{2}\,\phi_2^\dagger
\nonumber\\[7pt]
&-&
\mu^2\,\vert 1-\,\mu\,\nu_1\,\nu_2\,\xi\,\vert^2\,
\Big[\,\phi_1^\dagger\phi_1+\phi_2\,\phi_2^\dagger
\Big]
-\mu\,\Big[\,(1-\mu\,\nu_1\,\nu_2\,\xi\,)\,\,\psi_1\,\psi_2 + h.c.\Big]
\nonumber\\[7pt]
&-&
2\,\xi'\,\mu\,\Big[(\phi_1\phi_2)+h.c.\Big]\,
\Big[\,\phi_1^\dagger\phi_1+\phi_2\,\phi_2^\dagger
\,\Big],\qquad \cD^2=\cD^\mu\,\cD_\mu
\eea

\medskip
\noindent
This Lagrangian is in agreement with that of (\ref{adasd}).
This shows that onshell and in the absence of other interactions,
only a wavefunction renormalisation effect is present,
giving a  new  $\mu'= \mu\,\,(1-\mu\,\nu_1\,\nu_2\,\xi)$.
To conclude, integrating out the massive superfields $\Phi_{3,4}$
generated  a dimension-five operator $\Phi_2\,e^{-V}
D^2\,e^V\,\Phi_1$,
which if set
onshell via equations of motion or using the offshell Lagrangian,
brings  a (classical) wavefunction renormalisation only, in
 the absence of additional trilinear
interactions.  Thus this D=5 operator
does not bring new physics of its own, in the absence of
additional interactions.  One can then ask  whether this
conclusion remains true\footnote{without setting onshell
this operator}  after  supersymmetry is softly
broken, and this is  answered in the text,  Section~\ref{ss1} and
\ref{ss2}. To this purpose
the  supersymmetry  breaking terms associated to
this dimension-five operator must firstly be identified,
and this is done in  Appendix~\ref{appendixC}.
Finally, if additional, trilinear interactions 
were also present, other  dimension-five operators of type 
shown in (\ref{B7})  could also generated
 and these were also analysed  in Section~\ref{ss1}.

\def\theequation{C-\arabic{equation}}
\def\thesubsection{C}
\setcounter{equation}{0}

\subsection{Supersymmetry breaking effects and  higher
  dimensional operators.}
\label{appendixC}

In this appendix we  find all the
supersymmetry breaking  terms associated with the higher dimensional
operator  $\Phi_2\, e^{-V}\,D^2\,e^{V} \Phi_1$, which were
used in  Section~\ref{ss1} and \ref{ss2}. This operator is generated
as shown in (\ref{ff2})  (no gauge interactions) and  in
(\ref{lag4}) by integrating out massive superfields\footnote{
It would be more appropriate to introduce supersymmetry breaking to
$\cL_2$ of (\ref{LO}) then integrate again $\Phi_{3,4}$.
It is however easier to start from (\ref{lag4}) and add
to that a general spurion dependence/susy breaking.}.
 To find its associated  susy breaking contribution
 we use the spurion field technique and
 claim that the most general susy breaking
terms coming from this operator are generated by:
\medskip
\bea\label{lgs}
\cL_{G,S}&\equiv& \frac{1}{M}\int d^{4}\theta
\,\,\,A(S,S^\dag)\,D^{\alpha}\,\Big[B(S,S^\dag)\,\Phi_2e^{-V}
\Big]\, D_{\alpha}\,\Big[\Gamma(S,S^\dag)\,e^{V}\Phi_1 \Big]
 +h.c.
\eea
\medskip\noindent
where
\bea\label{definitions}
A(S,S^\dagger)&=& \alpha_0+\alpha_1\,S+\alpha_2\,S^\dagger+
\alpha_3\,S\,S^\dagger\nonumber\\[8pt]
B(S,S^\dagger)&=&\beta_0+\beta_1\,S+\beta_2\,S^\dagger+
\beta_3\,S\,S^\dagger\nonumber\\[8pt]
\Gamma(S,S^\dagger)&=&
\gamma_0+\gamma_1\,S+\gamma_2\,S^\dagger+\gamma_3\,S\,S^\dagger
\eea

\medskip\noindent
$A, B, \Gamma$ are the most general
spurion fields, and $S=\theta^2 M_s$, where $M_s$ denotes
the scale of
supersymmetry breaking. Also
$\alpha_i, \beta_i,\gamma_i$ are arbitrary input parameters of the theory.
 In (\ref{lgs}) an overall factor from spurion superfields
can always be absorbed into a redefinition of the scale M.
This is equivalent to saying that
$\alpha_0, \beta_0, \gamma_0$ can be set to unity.
However,  these can also vanish, therefore we kept their presence
explicit.
After a long calculation one finds
\bigskip
\bea\label{G}
\cL_{G,S}&=& -
\frac{\alpha_0\beta_0\gamma_0}{M}
 \int d^4\theta\,\,\Phi_2\,e^{- V}
 \, D^2 \,\big[e^{ V}\,\Phi_1\big]
\nonumber\\[10pt]
&+&
\frac{M_{s}}{M}
\big[4(d_1+d_2)\,\, \phi_2 \cD^\mu\cD_{\mu}\phi_1
-2\,(d_1-d_2)\,\,
\phi_2 D\phi_1  +2\sqrt{2}
\,d_1\,\,\phi_2 \lambda \psi_1 \nonumber \\[10pt]
&-&2\sqrt{2}\,d_2\,\,\psi_2\lambda \phi_1  -4d_3\,\,F_2F_1\big] +
\frac{M_{s}^2}{M}\big[-4d_4\,\,\phi_2 F_1 -4d_5\,\,F_2\phi_1
+2d_6\,\,\psi_2\psi_1\big] \nonumber \\[10pt]
&+&\frac{M_{s}^3}{M}\big[-4d_7\,\, \phi_2 \phi_1 \big] +h.c.
\eea

\bigskip\noindent
where the exact susy term can be read from the last three lines
of (\ref{offshell}) proportional to $\xi$, and
$h.c.$ applies to all terms;
the coefficients $d_i$, $i=1,7$ are given by:
\medskip
\bea\label{dis}
d_1&=& - \beta_1\, \alpha_0\,\gamma_0\,
-\frac{1}{2}\,\alpha_1\,\beta_0\,\gamma_0,
\qquad\qquad\qquad
d_2 = -\gamma_1\,\beta_0\,\alpha_0
-\frac{1}{2}\,\alpha_1\,\beta_0\,\gamma_0 \nonumber
\\
d_3 &= &-\alpha_2\,\beta_0\,\gamma_0
-\alpha_0\beta_2\gamma_0-\alpha_0\beta_0\gamma_2, 
\qquad\quad
d_4= -\beta_3\,\alpha_0\,\gamma_0-\beta_1\,\alpha_2\,\gamma_0
-\alpha_0\beta_1\gamma_2
\eea
and
\bea\label{dis2}
d_5& = & -\gamma_3\,\beta_0\,\alpha_0-\gamma_1\,\alpha_2\,\beta_0
-\alpha_0\beta_2\gamma_1,
\qquad\,\,
d_6 = \alpha_3\,\gamma_0\,\beta_0
+\alpha_1\beta_2\gamma_0
+\alpha_1\beta_0\gamma_2
\nonumber \\
d_7&=& -\gamma_3\,\beta_1\,\alpha_0-\gamma_1\,\beta_3\,\alpha_0-
\gamma_1\,\beta_1\,\alpha_2.
\eea

\medskip\noindent
Note the presence of the term $\phi_2\,D\,\phi_1$
(assuming $d_1-d_2\not=0$), where $D$ is the auxiliary gauge field.
This term and  $\psi_2\lambda\phi_1$
are not present in the MSSM, if we replaced
$\Phi_{1,2}$ by the MSSM Higgs fields $H_{1,2}$.

\bigskip
\def\thesubsection{D}
\def\theequation{D-\arabic{equation}}
\setcounter{equation}{0}
\subsection{Mass eigenvalues
in the MSSM with higher dimensional operators.}\label{appendixD}

Some details of the calculation in Section~\ref{ss3}
are given below.
From the two minimum conditions for the scalar potential $V$ of
 eq.(\ref{scalarV})
one can  express $\tilde m_{1,2}$ there in terms
of $B\mu$, $  v_1,   v_2$ to find:
\bea\label{tildemi}
\tilde m_1^2 &=& {-B\mu}\,\,\frac{  v_2}{  v_1}
-\frac{1}{8}\,g^2\,(  v_1^2-  v_2^2)
-\frac{\eta_1}{2}\,\frac{  v_2}{  v_1}\,
(3\,  v_1^2-  v_2^2)-\frac{\eta_2}{2}
\frac{  v_2}{  v_1}\,(3\,  v_1^2+  v_2^2)
-\frac{\eta_3}{2}\,{  v_2}^2 \nonumber\\[10pt]
\tilde m_2^2 &=&
 {-B\mu}\,\,\frac{  v_1}{  v_2}+\frac{1}{8}\,g^2\,({
   v_1}^2-{  v_2}^2)
-\frac{\eta_1}{2}\,\frac{  v_1}{  v_2}\,(
v_1^2-3\,  v_2^2)
-\frac{\eta_2}{2}\,\frac{  v_1}{  v_2}
\,(3\,  v_2^2+  v_1^2)-\frac{\eta_3}{2}\,  v_1^2\qquad
\eea

\medskip\noindent
which shall be used in the following.
The mass matrix is

\bea\label{mij}
\cM_{ij}&=&
\frac{1}{2}\frac{\partial^2 V}{\partial h_i\partial
  h_j}\bigg\vert_{h_i=  v_i/\sqrt{2},\,\tilde\sigma_i=0}
 =X_{ij}+Z_{ij}
\eea
where
\bea
X_{ij}=\frac{1}{2}
\left(
\begin{array}{cc}
2\tilde m_1^2+\frac{1}{4} \,g^2\,(3 v_1^2-v_2^2) &  2 B\,\mu-\frac{1}{2}\,g^2
v_1 \,v_2\\[12pt]
2 \,B\,\mu-\frac{1}{2}\,g^2\,v_1 \,v_2  &  2\tilde m_2^2
-\frac{1}{4}\,g^2\,(v_1^2-3\,v_2^2)
\end{array}
\right)
\eea
and
\bea
Z_{ij}=\frac{1}{2}
\left(
\begin{array}{cc}
6\,(\eta_1+\eta_2)\,v_1\,v_2+\eta_3\,v_2^2  &
\!\!\!\!\!\!\!\!\!\!\!\!\!\!\!\!\!
3\,(\eta_1+\eta_2)\,v_1^2
+3 (\eta_2-\eta_1)\,v_2^2+2 \eta_3\,v_1\,v_2
\\[12pt]
3\,(\eta_1+\eta_2)\,v_1^2
+3 (\eta_2-\eta_1)\,v_2^2+2 \eta_3\,v_1\,v_2 &
\!\!\!\!\!\!\!\!\!\!\!\!\!\!\!\!\!
6\,(\eta_2-\eta_1)\,v_1\,v_2+\eta_3\,v_1^2
\end{array}
\right)
\eea

\bigskip
\noindent
The mass eigenvalues  $m_{h,H}^2$ of
$\cM_{ij}$ are

\bea\label{difference}
m_{h, H}^2&=&M_{h, H}^{2}
\mp
 \frac{6\eta_1}{\sqrt w}
\bigg[B\mu\,(  v_1^2-  v_2^2)+  v_1   v_2
\Big(\tilde m_1^2-\tilde
  m_2^2+\frac{g^2}{4}(  v_1^2-  v_2^2) \Big)\bigg]
\nonumber\\[10pt]
&+&3\eta_2\,\bigg[  v_1   v_2
\pm\frac{1}{2\sqrt w}(  v_1^2+  v_2^2)(-4B\mu
+ g^2 \,   v_1   v_2)\bigg]
\nonumber\\[10pt]
&+&
\frac{\eta_3}{4}\bigg[
  v_1^2+  v_2^2
\pm \frac{1}{\sqrt w}
\Big(2 (\tilde m_1^2-\tilde m_2^2)(  v_1^2-
v_2^2)+g^2(  v_1^2+  v_2^2)^2-16
B \mu   v_1\,   v_2\Big)\bigg]
\eea

\bigskip\noindent
The upper (lower)
signs correspond to the lighter $ m_h^2$
(heavier $ m_H^2$) Higgs
field, respectively.
We introduced
\bea
M_{h,H}^{2}\equiv
\frac{1}{2}\bigg[\tilde m_1^2+\tilde
  m_2^2+
\frac{g^2}{4}\,(  v_1^2+  v_2^2)\mp \frac{1}{2}\sqrt w\bigg]
\eea

\medskip
\noindent
where the upper (lower) sign corresponds to
$M_h$ ($M_H$) which, if $\eta_{1,2,3}=0$ reproduce
the lighter (heavier)  MSSM Higgs field.
Above we used the notation
\medskip
\bea\label{rw}
w& \equiv &
(4 B\mu-g^2   v_1   v_2)^2+4\Big(\tilde m_1^2-\tilde m_2^2+
\frac{g^2}{2}(  v_1^2-  v_2^2)\Big)^2 
\eea

\noindent
With the values of $\tilde m_{1,2}$ expressed in terms of
$v_{1,2}$ and $B\mu$ from minimum conditions (\ref{tildemi}),
one can re-express  $ m_{h,H}^2$ of (\ref{difference})
as follows

\bea\label{mhH}
 m_{h, H}^2&=&\frac{m_Z^2}{2}
-\frac{B\mu (u^2+1)}{2\,u} \mp\frac{\sqrt w'}{2}
+{  v}^2\,\Big[\eta_1\,\,q_1^{\pm}
+\,\eta_2\,\,q_2^{\pm}
+\,\eta_3\,\,q_3^{\pm}\Big]
\eea
with
\bea
\! q_1^\pm & = & \frac{u^2-1}{4\,u}\pm
\frac{(u^2-1)}{4 u^2(1+u^2)^2\sqrt w'}\,\Big[
m_Z^2\,u (1-6 u^2+u^4)+ B\mu\,(1+u^2)(1+18 u^2+u^4)\Big]
\nonumber\\[8pt]
q_2^\pm & = &\!\!\!
- \frac{1-6u^2+u^4}{4\,u\,(1+u^2)}
\mp \frac{1}{4 \, u^2\,(1+u^2)\sqrt w'}
\Big[m_Z^2 u(1-14 u^2+u^4)+B\mu (1+u^2) (1+10 u^2 +u^4)
\nonumber\\[8pt]
q_3^\pm & = &
\mp \frac{2 u}{(1+u^2)^2\sqrt
  w'}\,\Big[B\mu(1+u^2)-m_Z^2\,u\Big]
\eea
where
\bea
w'\equiv m_Z^4 + \big[B\mu (1+u^2)^3+2 m_Z^2 u(1-6 u^2 +u^4)
\big]\frac{B\mu}{u^2(1+u^2)}
\eea

\medskip\noindent
and where we also used
$  v_1=  v\cos\beta,   v_2=  v\,\sin\beta$,
$u=\tan\beta$  and $m_Z^2=g^2\,{  v}^2/4$.
Similar considerations apply for the
pseudoscalar Higgs/Goldstone boson sector. The mass matrix is
in this case
\medskip
\bea\label{nij}
N_{ij}&=&\frac{\partial^2 V}{\partial
  \tilde\sigma_i\partial\tilde\sigma_j}\bigg\vert_{
{ h_i=  v_i}/{\sqrt 2},\,\tilde\sigma_i=0}
\eea
with entries
\bea
N_{11}&=&
\tilde m_1^2+\frac{g^2}{8}\,(  v_1^2-  v_2^2)
+(\eta_1+\eta_2)  v_1  v_2 -\frac{\eta_3}{2}  v_2^2
\nonumber\\[8pt]
N_{12}&=&
-\frac{\eta_1}{2}(  v_1^2-  v_2^2)-\frac{\eta_2}{2}
(  v_1^2+  v_2^2)-\eta_3  v_1  v_2-Re(B\mu)
\nonumber\\[8pt]
N_{22}&=&
\tilde m_2^2-\frac{g^2}{8}\,(  v_1^2-  v_2^2)+
(\eta_2-\eta_1)  v_1  v_2 -\frac{\eta_3}{2}  v_1^2
\eea

\noindent
The eigenvalues of $N$ are
\medskip
\bea\label{mga}
m_{G, A}^2&=&
\frac{1}{2}\,\big(\tilde m_1^2+\tilde m_2^2)\mp
\frac{1}{8}\sqrt \kappa
\nonumber\\[10pt]
&\mp&
\frac{4 \eta_1}{\sqrt \kappa}\,
\Big[  B\mu (  v_1^2-  v_2^2)
+
   v_1\,    v_2\,  \Big(\tilde  m_1^2 - \tilde  m_2^2 +
\frac{g^2}{4} (  v_1^2 -   v_2^2)\Big)\Big]
+
\eta_2\,\Big[  v_1  v_2\mp \frac{4  B\mu}{\sqrt \kappa}
\,(  v_1^2+  v_2^2)\Big]
\nonumber\\[10pt]
&+&
\eta_3\,\Big[
-\frac{1}{4}\,(  v_1^2+  v_2^2)
\mp\frac{1}{\sqrt \kappa }\Big(
8 B\mu   v_1  v_2\, + (  v_1^2-  v_2^2)(\tilde
m_1^2- \tilde m_2^2)+\frac{g^2}{4} \, (  v_1^2-  v_2^2)^2\Big)
\Big]
\eea
where
\bea
\kappa=
16\Big[4 (B\mu)^2 + \Big(\tilde m_1^2 - \tilde m_2^2 +
  \frac{g^2}{4}\,(  v_1^2-  v_2^2)\Big)^2\Big]
\eea

\medskip\noindent
where the upper sign corresponds to the Goldstone $m_G$  and
the lower sign to $m_A^2$.
One can
use (\ref{tildemi}) to
replace $\tilde m_{1,2}$ in terms of  $  v_{1,2}$ and $m_A$ .
Using (\ref{tildemi}) one shows  that  $m_G=0$ and
\medskip
\bea\label{ma}
m_A^2&=& -\frac{  v_1^2+  v_2^2}{2   v_1   v_2}
\,\Big[\,2 \,B\,\mu+
\eta_1\,(  v_1^2-  v_2^2)
+
\eta_2\,(  v_1^2+  v_2^2)
+
2 \eta_3 \,  v_1\,  v_2\Big]
\nonumber\\[10pt]
&=&
-
\frac{1+u^2}{u}\,B\,\mu
+
\, \frac{u^2-1}{2 \,u}\, \eta_1 \,v^2\,
-
\,\frac{1+u^2}{2\,u}\,\eta_2\,v^2\,
-
\eta_3\,v^2
\eea

\medskip\noindent
This is the result  used in the text, eq.(\ref{maprime}).
Using eqs.(\ref{mhH}) and (\ref{ma}) to eliminate
$B\mu$ between them,  one obtains the masses
$m_{h,H}$:
\medskip
\bea\label{finalmh}
m_{h,H}^2&=&
\frac{1}{2}\Big[m_A^2+m_Z^2\mp\sqrt{w''}\Big]
\mp
\frac{4\,m_A^2\,\eta_1\,u\,(u^2-1)\,{  v}^2}{(1+u^2)^2\,\sqrt{w''}}
\nonumber\\[10pt]
&+&
\frac{2\eta_2\,u\,{  v}^2}{1+u^2}
\,\bigg[1\pm\frac{m_A^2+m_Z^2}{\sqrt{w''}}\bigg]+
\frac{\eta_3\,{  v}^2}{2}\,\bigg[1\mp
\frac{(m_A^2-m_Z^2)\,(u^2-1)^2}{\sqrt{w''}\,\,\,(1+u^2)^2\,}\bigg]
\eea

\medskip
\noindent
where the upper (lower) signs correspond to $h$ ($H$) respectively,
and where
\medskip
\bea\label{mhf}
w'' & \equiv &
 m_A^4+m_Z^4-2\,m_A^2\,m_Z^2\,\frac{1-6u^2+u^4}{(1+u^2)^2}
=
(m_A^2+m_Z^2)^2-4\,m_A^2\,m_Z^2\,\cos^2 2\beta\qquad
\eea

\medskip\noindent
Replacing $u=\tan\beta$ in  $m_{h,H}$ one obtains
an equivalent form of $m_{h,H}$
 used in the text,  eq.(\ref{mh2}).

The bounds on $\eta_i$ discussed in Section~\ref{ss3}
that must be respected
in order to increase $m_h>m_Z$ in the approximation
considered, are derived from (\ref{finalmh}) with  (\ref{cond3})
and give
\medskip
\bea\label{ssx3}
&&\qquad \frac{(\sqrt\omega+1-\rho)\,(1+u^2)^2\,\sqrt\omega}{
32 \,u\,(u^2-1)}
\leq  -\frac{\eta_1}{g^2}
\ll
\min\bigg\{\frac{u}{6(u^2-1)}, \frac{3\,u^2-1}{24 u},
\frac{\vert u^2-3\vert }{24\,u}
\bigg\}\nonumber\\[8pt]
&&\frac{(\sqrt\omega+1-\rho)\,(1+u^2)^2\,\sqrt\omega}{
4 \,[(1+u^2)^2 \sqrt \omega-(\rho-1)(1-u^2)^2]}
\leq \frac{\eta_3}{g^2}
\ll
\min\bigg\{\frac{1}{4},
\frac{\vert  u^2-3\vert}{4\,u^2},
\frac{u^2-1}{4\,u^2},\frac{u^2-1}{4}
\bigg\}\qquad\qquad
\eea

\medskip
\noindent
with $\omega\equiv (\rho-1)^2+16 u^2\rho/(1+u^2)^2$,
 $u\equiv \tan\beta$ and $\rho\equiv m_A^2/m_Z^2$.
The implications of these eqs
 are discussed in the text after eq.(\ref{cond3}).

\end{document}